\documentclass[twocolumn]{aastex63}

\usepackage{float}

\usepackage{amsmath}	

\usepackage[normalem]{ulem}

\newcommand{\tsim}[1]{\ensuremath{{\sim}#1}}

\newcommand*\rhocl{\ensuremath{\rho_{\rm cl}}}
\newcommand*\rhow{\ensuremath{\rho_{\rm w}}}
\newcommand*\nmix{\ensuremath{n_{\rm mix}}}
\newcommand*\ncl{\ensuremath{n_{\rm cl}}}
\newcommand*\nw{\ensuremath{n_{\rm w}}}
\newcommand*\Tmix{\ensuremath{T_{\rm mix}}}
\newcommand*\Tcl{\ensuremath{T_{\rm cl}}}
\newcommand*\Tw{\ensuremath{T_{\rm w}}}

\newcommand*\mucl{\ensuremath{\mu_{\rm cl}}}

\newcommand*\rcl{\ensuremath{R_{\rm cl}}}
\newcommand*\rclcrit{\ensuremath{R_{\rm cl,crit}}}
\newcommand*\vw{\ensuremath{v_{\rm w}}}

\newcommand*\Mw{\ensuremath{\mathcal{M}_{\rm w}}}
\newcommand*\tcc{\ensuremath{t_{\rm cc}}}

\newcommand*\tcool{\ensuremath{t_{\rm cool}}}
\newcommand*\tcoolmix{\ensuremath{t_{\rm cool, mix}}}
\newcommand*\tcoolcl{\ensuremath{t_{\rm cool, cl}}}
\newcommand*\tcoolw{\ensuremath{t_{\rm cool, w}}}
\newcommand*\tcoolclchar{\ensuremath{\tilde{t}_{\rm cool, cl}}}

\newcommand*\Kcl{\ensuremath{K_{\rm cl}}}
\newcommand*\Kw{\ensuremath{K_{\rm w}}}
\newcommand*\Kmix{\ensuremath{K_{\rm mix}}}
\newcommand*\Kdot{\ensuremath{\dot{K}}}
\newcommand*\Kdotcool{\ensuremath{\dot{K}_{\rm cool}}}
\newcommand*\Kdotmixing{\ensuremath{\dot{K}_{\rm mixing}}}
\newcommand*\Kdotmixingchar{\ensuremath{\dot{K}_{\rm mixing,char}}}
\newcommand*\Kdottotal{\ensuremath{\dot{K}_{\rm total}}}

\newcommand*\Kdotavg{\ensuremath{\dot{\bar{K}}}}
\newcommand*\Kdotavgmixing{\ensuremath{\Kdotavg_{\rm mixing}}}

\newcommand*\Kdotavgtotal{\ensuremath{\Kdotavg_{\rm total}}}

\newcommand*\Mps{\ensuremath{M_{\rm ps}}}
\newcommand*\dMpsdKinline{\ensuremath{d\Mps/dK}}
\newcommand*\dMpsdK{\ensuremath{\frac{d\Mps}{dK}}}

\newcommand*\lcool{\ensuremath{\ell_{\rm cool}}}

\newcommand*\plawratio{\ensuremath{\eta_{\rm mix,cl}}}

\newcommand*\simFid{\texttt{NR-X100}}
\newcommand*\simMachThreeQuarter{\texttt{NR-X100-M0.75}}
\newcommand*\simMachThree{\texttt{NR-X100-M3}}
\newcommand*\simMachFive{\texttt{NR-X100-M4.5}}

\newcommand*\simContrastTwo{\texttt{NR-X180}}
\newcommand*\simContrastThree{\texttt{NR-X300}}
\newcommand*\simContrastTen{\texttt{NR-X1000}}
\newcommand*\simBTen{\texttt{NR-X100-B10}}
\newcommand*\simBTHundred{\texttt{NR-X100-B100}}
\newcommand*\simBThousand{\texttt{NR-X100-B1000}}
\newcommand*\simRCSlowOne{\texttt{SlowRC-T1e4}}
\newcommand*\simRCFastFour{\texttt{FastRC-T4e4}}
\newcommand*\simRCFastOne{\texttt{FastRC-T1e4}}
\newcommand*\simBPLawOne{\texttt{BPLawRC-1}}
\newcommand*\simBPLawTwo{\texttt{BPLawRC-2}}
\newcommand*\simBPLawSix{\texttt{BPLawRC-6}}
\newcommand*\simBPLawSixty{\texttt{BPLawRC-60}}

\newcommand*\enzoe{{\sc enzo-e}}
\newcommand*\enzo{{\sc enzo}}
\newcommand*\cello{{\sc cello}}
\newcommand*\grackle{{\sc grackle}}

\received{?} 
\revised{?}  
\accepted{?} 
\submitjournal{ApJ}

\shorttitle{Mixing \& Cooling in Cloud-Wind Interactions}
\shortauthors{Abruzzo et al.}
\graphicspath{{./}{figures/}}

\begin{document}


\title{A simple model for mixing and cooling in cloud-wind interactions}

\correspondingauthor{Matthew W. Abruzzo}
\email{mwa2113@columbia.edu}


\author[0000-0002-7918-3086]{Matthew W. Abruzzo}
\affiliation{Department of Astronomy,
Columbia University,
New York, NY 10027, USA}

\author[0000-0003-2630-9228]{Greg L. Bryan}
\affiliation{Department of Astronomy,
Columbia University,
New York, NY 10027, USA}
\affiliation{Center for Computational Astrophysics,
Flatiron Institute, 162 5th Ave,
New York, NY 10003, USA}

\author[0000-0003-3806-8548]{Drummond B. Fielding}
\affiliation{Center for Computational Astrophysics,
Flatiron Institute, 162 5th Ave,
New York, NY 10003, USA}




\begin{abstract}
  We introduce a simple entropy-based formalism to characterize the role of mixing in pressure-balanced multiphase clouds, and demonstrate example applications using \enzoe\ (magneto)hydrodynamic simulations.
  Under this formalism, the high-dimensional description of the system's state at a given time is simplified to the joint distribution of mass over pressure ($P$) and entropy ($K=P\rho^{-\gamma}$). 
  As a result, this approach provides a way for (empirically and analytically) quantifying the impact of different initial conditions and sets of physics on the system evolution. 
  We find that mixing predominantly alters the distribution along the $K$ direction and illustrate how the formalism can be used to model mixing and cooling for fluid elements originating in the cloud. 
  We further confirm and generalize a previously suggested criterion for cloud growth in the presence of radiative cooling, and demonstrate that the shape of the cooling curve, particularly at the low temperature end, can play an important role in controlling condensation.
  Moreover, we discuss the capacity of our approach to generalize such a criterion to apply to additional sets of physics, and to build intuition for the impact of subtle higher order effects not directly addressed by the criterion.
\end{abstract}



\section{Introduction} \label{sec:intro}




Stellar feedback driven galactic outflows play a critical role in galaxy formation and evolution.
They are important for regulating star formation and transporting metals out of galaxies \citep{white78a, dekel86a, white91a}. 
To reproduce observed galaxy properties, large-scale cosmological simulations often assume that these galactic winds effectively accelerate cool gas, with mass loading factors of \tsim{1-100} \citep[e.g.][]{pillepich18a,dave19a}.
Moreover, observations of outflows serve as direct evidence for the existence of stellar feedback.
A robust test of simulations is their ability to reproduce these observations \citep{somerville15a}.

Observations have revealed that these winds are inherently multi-phase
\citep[for reviews of observations see][]{veilleux05a, rupke18a};
there is cool gas comoving with hot gas. This multi-phase nature
presents a challenge to the conventional model that the outflows are
driven by hot (${\ga}10^6$ K) winds produced by supernovae. 
There has been a considerable effort to determine whether ram pressure acceleration of cool ($\sim$$10^4$ K) ISM, by these winds, can produce a co-moving multi-phase flow \citep[e.g.][]{klein94a, cooper09a,
  scannapieco15a, schneider17a, sparre18a}.
Difficulties arise because hydrodynamical instabilities (e.g. Kelvin-Helmholtz and Rayleigh-Taylor) grow from the initial velocity difference between the cloud and wind, and drives mixing between the phases; the phases can be homogenized before the cloud is entrained.

\citet{klein94a} showed that this destruction of cool clouds is roughly characterized by the {\it cloud-crushing time scale}. 
For a non-radiative cloud with density \rhocl\ and radius \rcl , initially at rest with respect to a hot wind, with density $\rhow = \rhocl / \chi$ ($\chi\sim100$--$1000$) and velocity \vw , this time-scale is given by
\begin{equation}
  \label{eqn:tcc}
  \tcc = \chi^{1/2} \frac{\rcl}{\vw}.
\end{equation}
Because the cloud is destroyed within a few \tcc, and \tcc\ is a factor of
${\sim}\chi^{1/2}$ smaller than the 
ram pressure acceleration time-scale, it's challenging for hot winds to entrain the cool gas before it's destroyed.

Subsequent studies have modeled additional physical effects in attempts to delay cloud disruption long enough for them to be embedded within the wind. 
The most common additional set of physics is radiative cooling \citep[e.g.][]{cooper09a, scannapieco15a, schneider17a}.
However, the general consensus was that the cloud's lifetime is not prolonged enough for it to be fully entrained in the wind. 
Moreover, \citet{zhang17a} compellingly showed (with semi-analytic methods) that real observations cannot be reproduced by ram-pressure accelerated cold clouds in the interval of time before they are destroyed.

Another approach for extending the cloud's lifetime has been the inclusion of magnetic fields. 
Certain configurations can inhibit mixing 
and provide an additional tension force that resists destruction. 
\citet{mccourt15a} demonstrated that the presence of a tangled magnetic field in the cloud gave promising results for $\chi=50$ (with the inclusion of radiative cooling).
Unless magnetic pressure dominates ($\beta<1$) in the wind, however, magnetic fields alone don't appear to inhibit the disruption of higher density contrast clouds ($\chi\sim100$--$1000$)  enough to allow their entrainment \citep{gronke20a}.

Other models have also been proposed to produce multi-phase outflows without requiring thermal supernovae winds to entrain clouds.
Alternatives include the acceleration of outflowing gas by non-thermal feedback like radiation pressure \citep[e.g.][]{zhang18a} and cosmic rays \citep[e.g.][]{wiener19a} or {\it in situ} cloud formation within a cooling outflow \citep[e.g.][]{thompson15a,schneider18a,lochhaas20a}.
These alternatives have achieved varying degrees of success, but no single model appears to apply in all cases.

Recent work \citep{armillotta16a,gronke18a} has shed new light on cloud acceleration by a hot wind in the limit of rapid cooling.
\citet{gronke18a,gronke20a} showed that if mixed gas cools sufficiently fast, then it becomes a part of the colder cloud phase before it is further homogenized with the wind.
We hereafter refer to this process as turbulent radiative mixing layer (TRML, which we pronounce as ``turmoil'') entrainment.
This process not only inhibits the depletion of the cloud mass, but also transfers mass {\it and} momentum to it from the wind (similar to an inelastic collision).

\citet{gronke18a} argued that this process occurs when the mixing time-scale, \tsim{\tcc}, exceeds the cooling times-scale of the mixing layer, \tcoolmix; the mixing layer has a temperature $\Tmix\sim\sqrt{\Tcl\Tw}$ and number density $\nmix\sim\sqrt{\ncl\nw}$. 
They recast this criterion, $\tcc>\tcoolmix$, as a radius requirement for a spherical cloud.
Clouds should survive when their radius exceeds
\begin{equation}
  \label{eqn:rcrit}
  \rclcrit \sim \frac{\vw \tcoolmix}{\chi^{1/2}}\approx
  2\, {\rm pc} \frac{T_{\rm cl,4}^{5/2} \Mw}{P_3 \Lambda_{\rm mix,-21.4}}
  \frac{\chi}{100},
\end{equation}
where $T_{\rm cl,4}\equiv \Tcl/(10^4\, {\rm K})$,
\Mw\ is the mach number of the wind, $P_3\equiv nT/(10^3\, {\rm
  cm}^{-3}\, {\rm K})$, and $\Lambda_{\rm mix,-21.4} \equiv
\Lambda(\Tmix)/(10^{21.4}\, {\rm erg}\, {\rm cm}^3\, {\rm s}^{-1})$\footnote{
Following the arguments from \citet{begelman90a}, (when $\mu$ has $\rho$ and $p$ dependence) we find that $\Tmix \sim \sqrt{\chi}\mu_{\rm mix} \Tcl/\mu_{\rm cl}$, $n_{\rm mix} \sim \mu_{\rm cl} \ncl/(\sqrt{\chi}\mu_{\rm mix})$, and \rclcrit\ scales with $\mu_{\rm cl}^{-5/2}$.
  Using \grackle\ \citep{smith17a}, in tabulated mode, we find slightly different characteristic values for solar metallicity, $\chi=100$, $\Mw=1$, $T_{\rm cl,4} =1$, and $P_3=1$
  Under these conditions, $\mucl=0.827$  and $\rclcrit\sim6\,{\rm pc}$ ($\Lambda_{\rm mix}$ is unchanged).
}.

This picture has been bolstered by recent related studies of individual shear layers \citep{ji19a,fielding20a,tan20a}. These simulations lack the overall cloud geometry but are able to reach significantly higher resolution. The key finding from these studies is that the crucial parameter that determines the rate of cooling and the rate at which wind material is advected into the cloud is the ratio of the cooling time to the eddy turn over time. The eddy turn over time is comparable to the cloud crushing time.

Recently, \citealt{li20a} found a different survival criterion, based on the cooling time of the wind, \tcoolw. 
They argue that clouds survive, when $\tcoolw/\tcc<10\bar{f}$, while $\bar{f}$ is an order-unity term with weak power-law dependence (exponents range from \tsim{0.3} to \tsim{0.6}) on \rcl, \nw, and \vw.
\citet{sparre20a} reached a similar result (with additional \Mw\ dependence), while \citet{kanjilal20a} found results in support of the \citet{gronke18a} criterion.
Therefore, this disagreement over the survival criterion remains unresolved.

Prior  works  have  clearly  assembled  models  for  how  various  circumstances  and  physical  effects modify the cloud-wind interaction.  Unfortunately, simple characterizations  of  how  different  effects  influence the interaction are often incompatible with one other. The complex multidimensional nature of this process is a barrier to making comparable and composable characterizations. We, therefore, explore higher order characterization of the mixing and cooling processes in these systems in order to isolate the competing effects.

In this paper, we present a simple entropy-based formalism for characterizing how different physical effects affect mixing (and other destruction processes).
This mixing model builds on the premise that changes in the pressure-entropy ($p-K$) phase distribution broadly capture the cloud-wind interaction's evolution.
Our approach is particularly conducive for comparing the impact of radiative cooling against mixing.
Furthermore, it naturally complements the \citet{gronke18a,gronke20a} physical model for turbulent radiative mixing layer entrainment.

Our paper is organized as follows: in \S\ref{section:motivation} we motivate and describe the formalism, and in \S\ref{section:methods} we describe the numerical methods used to test it.
Videos of our simulations can be found at \url{http://matthewabruzzo.com/visualizations/}.
Subsequently in \S\ref{section:results-no-cool} and \S\ref{section:results-cool}, we describe our results from two example applications that demonstrate how the formalism both: (i) broadly captures the system's evolution and (ii) can quantitatively characterize how different conditions and physical effects modify the system's evolution.
These applications include a non-radiative parameter study (\S\ref{section:results-no-cool}) and a more detailed study involving radiative cooling (\S\ref{section:results-cool}).
Finally, we discuss the significance of our results in \S\ref{section:discussion} and summarize our conclusions in \S\ref{section:conclusion}.

\begin{figure*}
  \center
\includegraphics[width = 6.in]{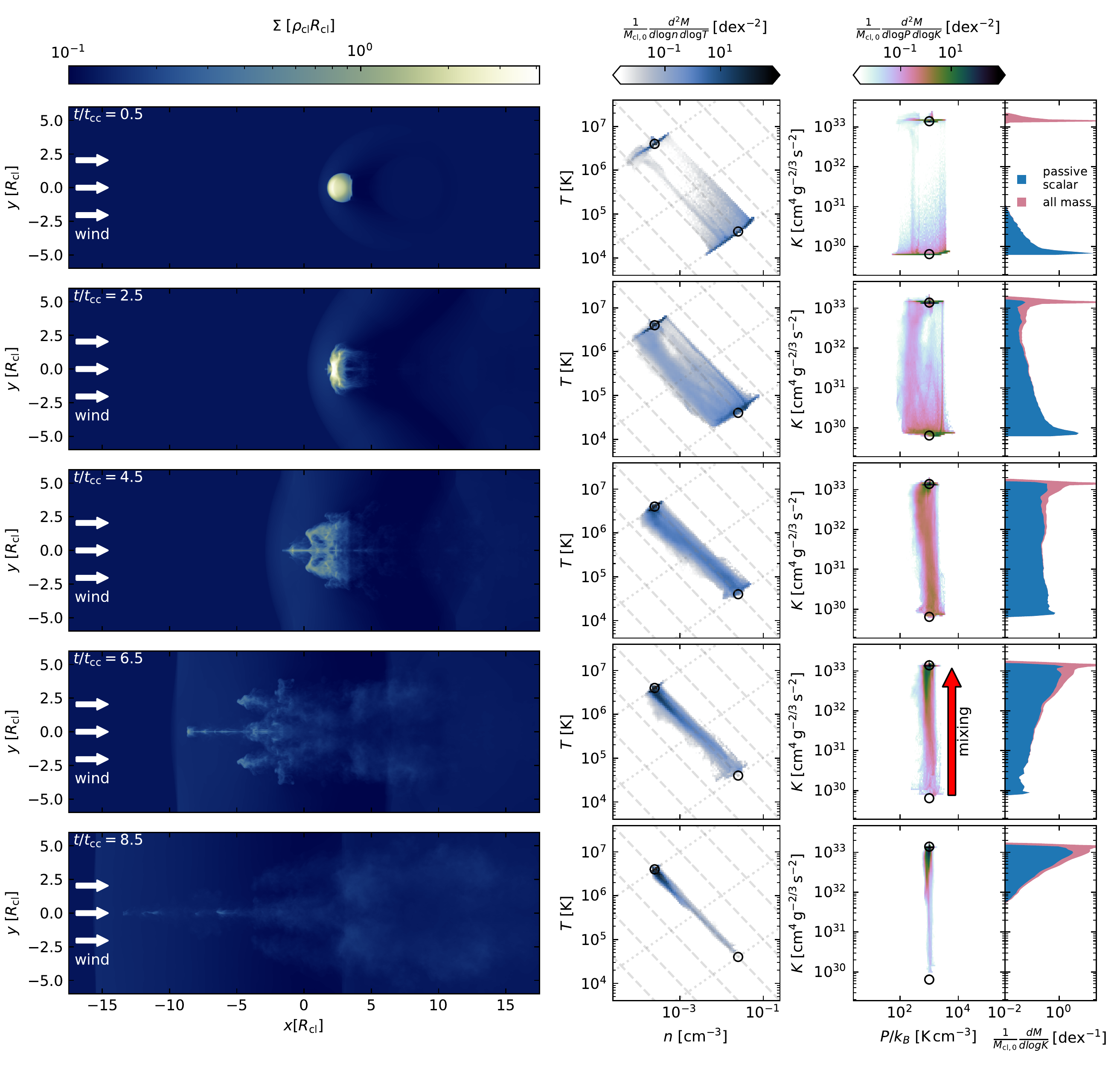}
\caption{\label{figure:example_evo} Illustration of the cloud-wind
  interaction until $8.5 \tcc$ for an initially spherical cloud, a
  wind with Mach number $\Mw=1.5$ and density contrast $\chi=100$ at a resolution of $\rcl/\Delta x = 64$.
  The left column illustrates the cloud surface density.
  We note that simulated domain is far longer than these panels depict (the shock remains in the domain over the entire simulation) and the axes labels don't describe the position in the domain.
  The center columns depict Mass-weighted $n-T$ (center-left) and $P-K$ (center-right) phase diagrams of {\it all} gas in the simulation.
  The $n-T$ diagrams are computed assuming a fixed mean molecular mass of $\mu=0.6$.
  The black circles near the top (bottom) of the panels denote the initial location of fluid elements originating in the wind (cloud).
  The dashed (dotted) gray lines in the $n-T$ diagram denote logarithmically spaced lines of constant $P$ ($K$). 
  The rightmost column depicts the one dimensional entropy distribution.
  The red histogram traces all gas in the domain, while the blue traces just the fluid elements initialized in the cloud.
}
\end{figure*}

\section{Model Overview}
\label{section:motivation}

\begin{figure}
  \center
\includegraphics[width = \columnwidth]{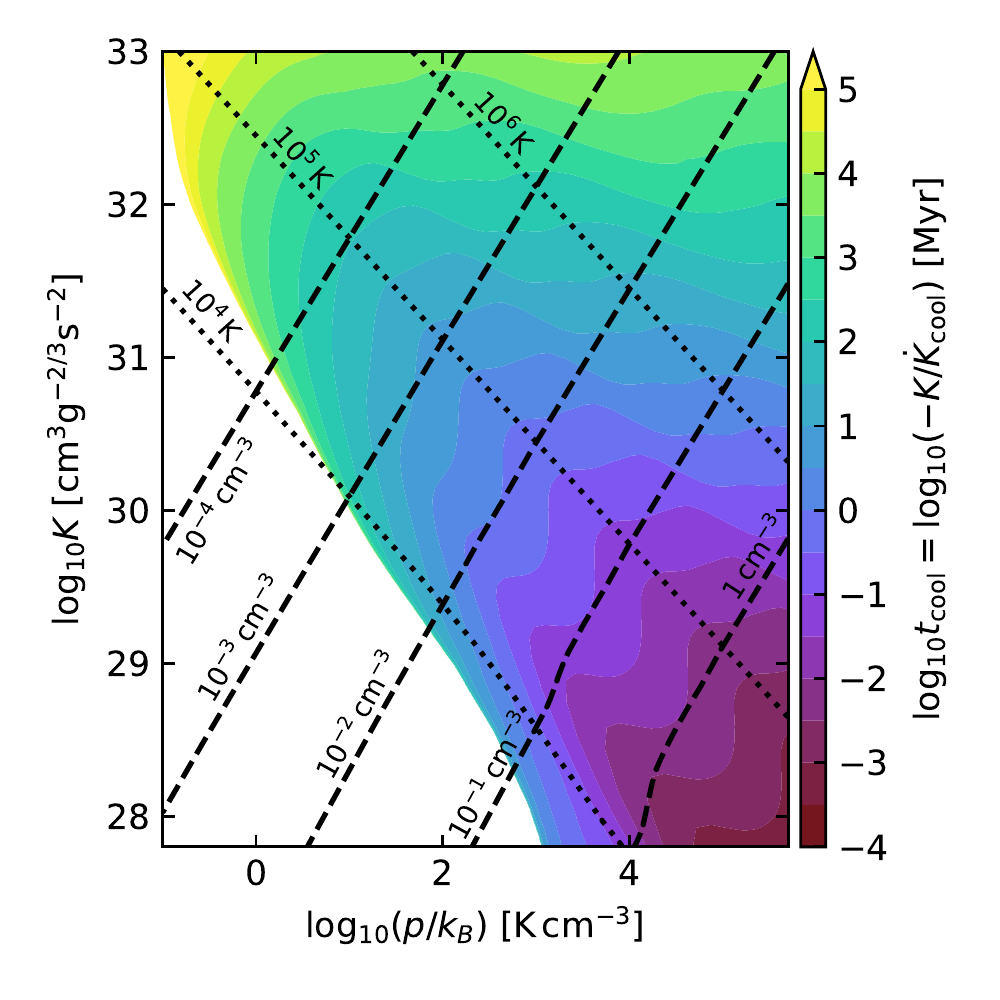}
\caption{\label{figure:tcool} Illustration of contours of
  $-K/\Kdotcool$ (equal to $t_{\rm cool}$ for a fixed composition,
  perfect gas) in pressure-entropy space.
  The dashed (dotted) denote logarithmically spaced number density (temperature) contours. 
  In the lower-left white region, heating dominates cooling.}
\end{figure}

\subsection{$p-K$ Representation}

The cloud-wind interaction is fundamentally an interaction between a finite pool of colder, dense gas (the cloud) and a large reservoir of hotter, more diffuse gas (the wind).
We are interested in understanding how different physical conditions affect the interaction's outcome.
The two outcomes are: (i) the homogenization of the colder phase within the more abundant hotter phase (cloud destruction) or (ii) the long-term survival and coexistence of both phases (entrainment).
It's instructive to specify the system's state purely in terms of this thermodynamic description.

Figure~\ref{figure:example_evo} helps illustrate this premise for a non-radiative simulation (see \S\ref{section:simulations} for more details about the $\rcl/\Delta =64$ \simFid\ simulation).
The left column illustrates snapshots of the system's morphological evolution while the center-left column shows the $n-T$ phase space evolution.
The initial properties of each gas phase are denoted by black circles in the phase diagrams.
At early times, the initial shock introduces pressure perturbations and slightly elevates the entropy of some fluid elements originating in the colder phase.
As the interaction progresses, adiabatic mixing drives gas from the colder dense phase towards the hotter, diffuse phase.
At a given snapshot, the phase distribution clearly encodes information about the system's state.

Although number density, $n$, and temperature, $T$, are familiar thermodynamic quantities, we choose to base our mixing model on $p-K$ phase space (center-right column of Figure~\ref{figure:example_evo}).
The quasi-isobaric nature of the problem makes pressure, $p$, an intuitive choice for a phase space axis.
Although the supersonic wind seeds small transient pressure perturbations, the second phase space dimension effectively indexes the continuum of properties gas can have between the two initial states.
Following convention, we pair $p$ with an entropy-like quantity, $K=p \rho^{-\gamma}$ (hereafter, we refer to $K$ as entropy) and we take $\gamma=5/3$ throughout this work.
The dashed (dotted) lines in the center column of Figure~\ref{figure:example_evo} denote contours of constant $p$ ($K$) that increase by factor of 10.

The choice of $p-K$ space over $p-T$ space is somewhat discretionary.
Because $T$ directly characterizes thermal energy, it more directly governs heat flow (e.g. cooling).
However, it's easier to characterize scale-free, dynamical effects in terms of $K$; relating $T$ to hydrodynamic quantities requires scale-dependent knowledge about the mean molecular mass, $\mu$.
Additionally, the entropy of a fluid element is unchanged by compression or expansion, which explains why the spread in gas at a given pressure lies along $K$ contours in the center column of Figure~\ref{figure:example_evo}.
Only irreversible processes change entropy: shocks and mixing increase it while radiative cooling decreases it.
Moreover, fluid elements have continuous trajectories through $p-K$ space in the absence of shocks because mixing and cooling modify $K$ smoothly.

We now take a more careful look at the $p-K$ representation and evolution for the non-radiative, hydrodynamic cloud-wind interaction.
At initialization, all gas has a single pressure $p$, the colder phase lies at $\Kcl=p \rho_{\rm cl}^{-5/3}$, and the hotter phase lies at $\Kw=\chi^{5/3}\Kcl$.
As the system evolves, hydrodynamical instabilities drive mixing of the two phases.
Mixing increases the entropy of the colder phase gas and initially decreases the entropy of the hotter phase gas.
By the time the cloud is destroyed, all gas has an entropy of $K\sim\Kw$.
The gas pressure remains relatively constant throughout this process.

The precise $p-K$ evolution depends on the initial conditions and the simulated physics.
Changes to either may modify how the distribution evolves.
Herein lies the true value of this thermodynamic description of the system's state: it provides a low-dimensional domain for characterizing mixing that is well-suited for comparing different physical processes.

\subsection{Mixing Model}
\label{section:mixing_model}

Consider the motion of fluid elements originating in the colder phase through $p-K$ space; this is illustrated by the red arrow in Figure~\ref{figure:example_evo}.
We anticipate factors that inhibit mixing, like strong magnetic fields, to decelerate the rate that fluid elements increases in entropy.
Conversely, we expect factors that hasten mixing, such as larger values of \Mw, to accelerate that rate.
Thus, we can empirically model mixing by characterizing the fluid elements' motion through $p-K$ space.

For simplicity, we largely ignore pressure perturbations in the context of mixing.
Motion along the pressure dimension may be particularly relevant in cases with strong sources of non-thermal pressure support.
The right column of Figure~\ref{figure:example_evo} illustrates just the entropy evolution of fluid elements.

In this work, we focus on the motion of fluid elements that originate in the cloud.
Because we trace these fluid elements with a passive scalar, we refer to their total mass as \Mps.
We defer analysis of the $p-K$ evolution for fluid elements originating in the hot phase to a future work.
These fluid elements encode information that is most relevant at times and locations in phase space where the motion of fluid element from the clouds aren't representative of all fluid elements at that location.
We briefly revisit this point in \S\ref{section:Kdot_method}.

Under these assumptions, the cloud-wind interaction at a time $t$ can be quantitatively described by its initial pressure ($p_0$), the distribution of the initial cloud fluid elements with respect to entropy, $(\dMpsdKinline)(K,t)$, and $\Kdot(K, t)$, the ensemble averaged Lagrangian derivative of $K$ for all initial cloud fluid elements with a given value of $K$.
The form of $\Kdot(K, t)$ dictates the outcome of the interaction.
It's dependent on the initial conditions and modeled physics (in full generality, the notation resembles $\Kdot(K, t; p_0, \gamma, \chi, \rcl, \Mw, \beta, \ldots)$).

For purely non-radiative (magneto)hydrodynamic interactions, the only sources of entropy are the initial shock and mixing.
Because we expect mixing to dominate outside of highly supersonic flows, we refer to $\Kdot(K, t)$ as $\Kdotmixing(K, t)$ for these simulations.
Because of the scale-free nature of such an interaction, the value of $\Kdotmixing(a \Kcl, b \tcc)/(\Kcl\tcc^{-1})$, where $a$ and $b$ are arbitrary positive values, is constant for any choice of \rhocl, $p_0$, and \rcl\ (as long as $\gamma$, $\chi$, $\Mw$,$\beta$, and the initial geometry remain unchanged).

When $\Kdotmixing(K,t)$ is known, it can be used to predict the outcome of interactions involving radiative cooling through comparisons against expected contributions from cooling, $\Kdotcool(p,K)$.
For optically thin gas, $\Kdotcool=K \dot{e}/e = -K/\tcool$ and Figure~\ref{figure:tcool} illustrates $\tcool(p,K)$.
When  $\Kdotmixing(K,t) + \Kdotcool(p_0,K) \gg 0$ for $K\in [\Kcl,\Kw]$, we generally expect the colder phase to be destroyed.
Conversely, the existence of a large sub-interval over $[\Kcl,\Kw]$, where the sum is far less than zero, suggests long term survival of a cool phase.
The outcome is ambiguous when the sum is close to zero because the sum does not directly give the total $\Kdot(K,t)$ for an interaction with cooling (hereafter $\Kdottotal(K,t)$).

This reasoning is reminiscent of the comparisons between \tcc\ and \tcoolmix\ that underlies the \citet{gronke18a} survival condition.
In fact, we can apply analogous arguments in the context of our mixing model to derive a criterion comparable to $\tcc > \tcoolmix$.
For simplicity, suppose $\Kdotmixingchar(K)$ gives the characteristic mixing rate for a non-radiative interaction at entropy $K$ at times when $\dMpsdKinline(K,t)>0$ (in \S\ref{section:mixing_rate} we confirm that $\Kdotmixingchar(K)$ is indeed a well-conceived quantity).
Then, following their logic, we expect turbulent radiative mixing layer entrainment to occur when $\Kdotmixingchar (\Kmix) < \left|\Kdotcool(p_0,\Kmix)\right|$, where the entropy at the mixing layer is $\Kmix\sim\sqrt{\Kcl\Kw}$.


\section{Methods}
\label{section:methods}
\subsection{Simulations}
\label{section:simulations}

\begin{table*}
	\centering
	\caption{Table of simulations (additional simulations are discussed in the appendices).
          All simulations were initialized with an initial thermal pressure of $p/k_B = 10^3\, {\rm cm}^{-3}\, {\rm K}$. 
          }
	\label{tab:sims}
	\begin{tabular}{ccccccccccc} 
	  \hline
          
          name & $\chi$    & $\Mw$\tablenotemark{a}    & $\beta$\tablenotemark{b}  & \rcl\ (pc) &
          \Tcl\tablenotemark{c} (K) & $R_{\rm cl}/\Delta x$ & Cooling\tablenotemark{d} & $\tcoolmix/\tcc$  &
          $\rcl/\lcool$\tablenotemark{e}  & $\tcoolmix/\tcoolcl$\\
	  \hline
	      \simFid               & 100   & 1.5  & $\infty$ & 1     & $4\times10^4$  & 8, 64 & N   &          &       & \\
          \simContrastThree     & 300   & 1.5  & $\infty$ & 1     & $4\times10^4$  & 8     & N   &          &       & \\
          \simContrastTen       & 1000  & 1.5  & $\infty$ & 1     & $4\times10^4$  & 8     & N   &          &       & \\
          \simMachThreeQuarter  & 100   & 0.75 & $\infty$ & 1     & $4\times10^4$  & 8     & N   &          &       & \\
          \simMachThree         & 100   & 3    & $\infty$ & 1     & $4\times10^4$  & 8     & N   &          &       & \\
          \simMachFive          & 100   & 3    & $\infty$ & 1     & $4\times10^4$  & 8     & N   &          &       & \\
	      \simBTen              & 100   & 1.5  & 10       & 1     & $4\times10^4$  & 8     & N   &          &       & \\
          \simBTHundred         & 100   & 1.5  & 100      & 1     & $4\times10^4$  & 8     & N   &          &       & \\
          \simBThousand         & 100   & 1.5  & 1000     & 1     & $4\times10^4$  & 8     & N   &          &       & \\
          \simRCSlowOne         & 100   & 1.5  & $\infty$ & 1     & $10^4$ & 64    & Y   & 8.982    & 0.668 & 1.044 \\
          \simRCFastFour        & 100   & 1.5  & $\infty$ & 5050  & $4\times10^4$  & 64    & Y   & 0.2165   & 388   & 63.06 \\
          \simRCFastOne         & 100   & 1.5  & $\infty$ & 57    & $10^4$ & 64    & Y   & 0.1576   & 38.1  & 1.044 \\
          \simBPLawOne          & 100   & 1.5  & $\infty$ & 66.51 & $10^4$ & 8     & BPL & 0.1576   & 9.52  & 1  \\
          \simBPLawTwo          & 100   & 1.5  & $\infty$ & 66.51 & $10^4$ & 8     & BPL & 0.1576   & 19.0  & 2  \\
          \simBPLawSix          & 100   & 1.5  & $\infty$ & 66.51 & $10^4$ & 8     & BPL & 0.1576   & 57.1  & 6  \\
          \simBPLawSixty        & 100   & 1.5  & $\infty$ & 66.51 & $10^4$ & 8     & BPL & 0.1576   & 571   & 60 \\
	  \hline
	\end{tabular}
        \tablenotetext{a}{sonic Mach number of the wind}
        \tablenotetext{b}{plasma beta (thermal pressure divided by magnetic pressure)}
        \tablenotetext{c}{For non-radiative simulations, \Tcl\ is computed assuming $\mu=0.6$.}
        \tablenotetext{d}{Indicates whether  radiative cooling is included. 
        Simulations with the value ``BPL'' used a custom broken power-law cooling curve with shaped given by the specified $\plawratio=\tcoolmix/\tcoolcl$ and Equation~\ref{eqn:broken_plaw} (and a constant $\mu$ of 0.6).}
        \tablenotetext{e}{\lcool\ is the ``cooling length,'' $\lcool=\min (\tcool c_s)$ \citep{mccourt18a}} 
\end{table*}

\begin{figure*}
  \center
\includegraphics[width = \textwidth]{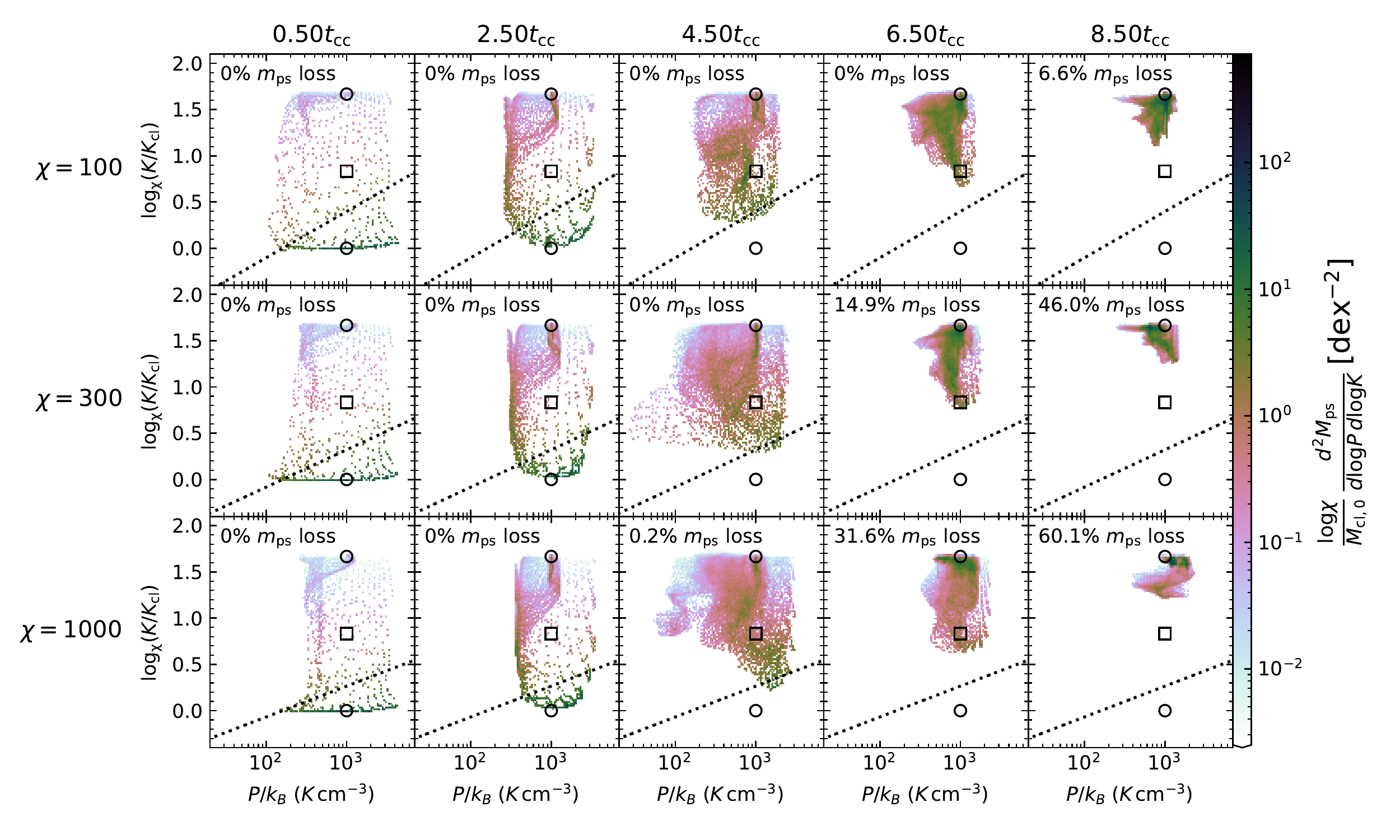}
\caption{\label{figure:p-K_phase_hydro}
  Mass-weighted Pressure-entropy phase evolution for gas originating in the cloud (in contrast Figure ~\ref{figure:example_evo} shows all gas in the simulation) for three non-radiative, hydrodynamic simulations.
  Each row depicts a simulation with a different $\chi$ (density contrast) and simulation time increases from left to right.
  The fraction of the gas originating in the cloud (i.e. the passive scalar) that has exited the simulation domain is recorded in each panel.
  We scale the entropy axes in terms of $\log_\chi (K/\Kcl)$ to facilitate easier comparisons between simulations, since the colder phase gas is initialized with
  $K=\Kcl$ and will have $K\sim\Kw=\chi^{5/3}\Kcl$ when it homogenizes with the hotter gas (these locations are indicated by black circles).
  Finally, the black square denotes $\Kmix=\sqrt{\Kcl,\Kw}$ and the dashed line line denotes $\rhocl/3$ (a common threshold used to identify cloud gas).
  }
\end{figure*}

\begin{figure*}
  \center
\includegraphics[width = \textwidth]{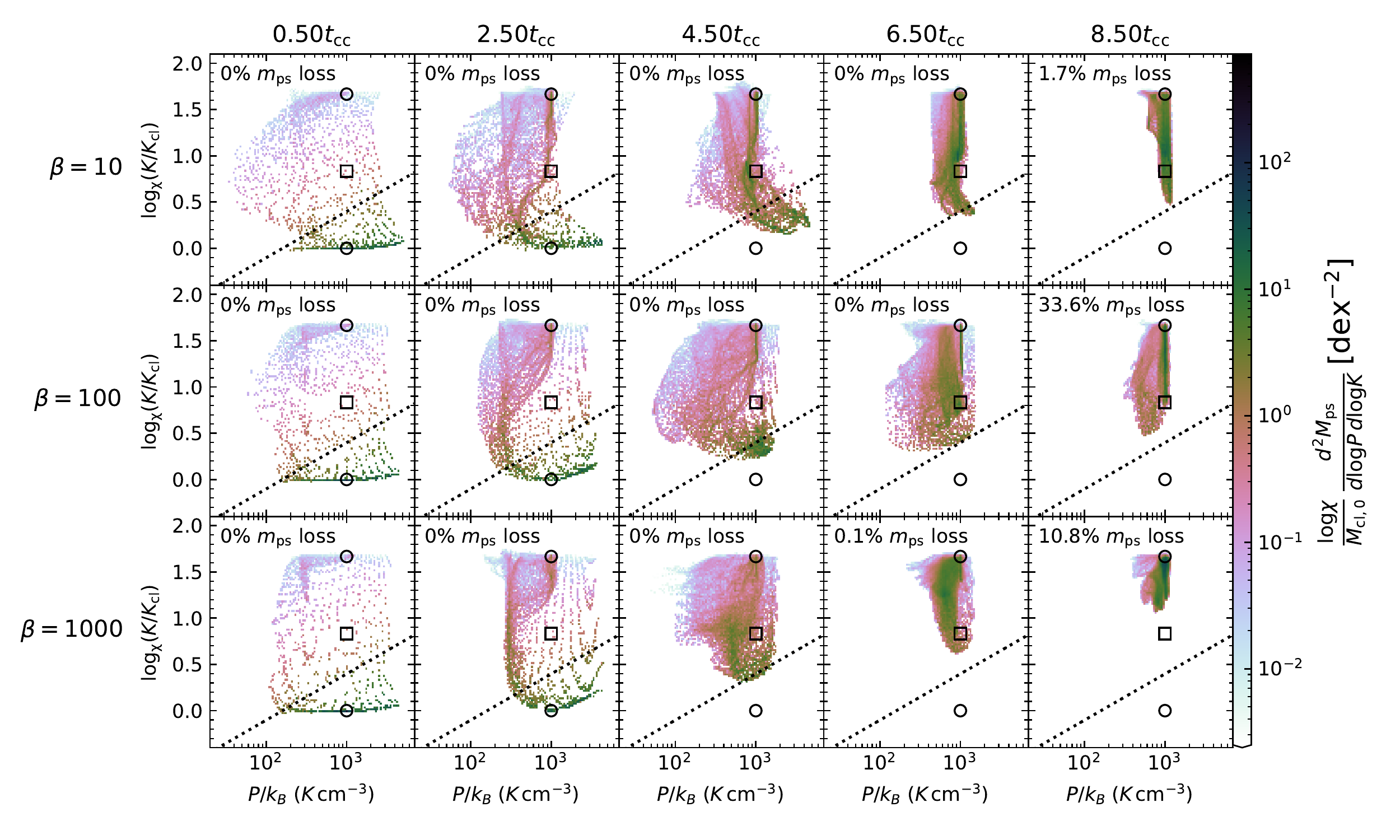}
\caption{\label{figure:p-K_phase_mhd} Mass weighted
  pressure-entropy evolution (for gas originating in the cloud) as in
  Figure~\ref{figure:p-K_phase_hydro} except that $\chi$ is fixed
  at 100 and the simulations include transverse magnetic fields.
  In each of the simulations, the initial magnetic fields is
  initialized with constant $\beta$ throughout the entire domain.}
\end{figure*}

To run magnetohydrodynamical (MHD) simulations for this work, we make
use of the \enzoe \footnote{\url{http://cello-project.org}} code.
This code is a rewrite of \enzo\ \citep{bryan14a} that targets exascale
computing and is built-on the distributed, scalable, adaptive mesh
refinement (AMR) framework, \cello\ \citep{bordner12a, bordner18a}.
Although \enzoe\ is still under active development, it
has matured enough that it can be used for basic scientific
studies.

For this work we implemented the second-order accurate unsplit VL + CT (van
Leer + Constrained Transport) algorithm presented by
\citet{stone09a}.
This is a predictor-corrector scheme
that employs the Constrained Transport (CT) method \citep{evans88a}.
Each simulation uses second order reconstruction\footnote{The prediction step always uses first order reconstruction} and the HLLD Approximate Riemann Solver \citep{miyoshi05a}.
We provide a brief assessment on how the choice of integrator affects the evolution of the cloud-wind interaction in Appendix~\ref{appendix:extra_survival_plots}.

In our cloud-wind simulations, we solve the ideal,
adiabatic MHD equations on a fixed, uniform three-dimensional
Cartesian grid. We initialize each simulation
with a spherical cloud of radius \rcl\  embedded in a steady
wind with $\Mw=1.5$ and a mass density that is a factor of
$\chi$ lower than the cloud. The cloud initially has no bulk
velocity and is in pressure equilibrium with the wind. We also
initialize a passively advected scalar that traces the gas
initially confined to the cloud.
In a subset of our non-radiative simulations we also initialize magnetic fields transverse to the wind that are constant throughout the entire domain.

Our runs including radiative cooling employ the \grackle \footnote{\url{https://grackle.readthedocs.io/}} chemistry and cooling library \citep{smith17a} and assume solar metallicity.
Our main cooling runs use the \citet{haardt12a} UV background model and don't use the self-shielding approximation.
In \S\ref{section:custom_curves} we also consider cases with custom broken power-law cooling curves.

For simplicity, we restrict cooling to only occur between ${\sim}\Tcl$ and ${\sim}0.6\Tw$; we modified the tables to have $\Lambda/m_H^2 = 10^{-99}\, {\rm erg}\, {\rm cm}^3\, {\rm s}^{-1}\, {\rm g}^{-2}$ in the restricted regions\footnote{Due to an oversight, CMB Compton cooling occurs (and actually dominates) in the restricted regions for simulations without broken power-laws.
For these cases $t_{\rm cool, restrict}\ga 3.4 n_H/(n_e \mu) \times10^{19}\, {\rm s} \ga 10^4\tcc$.
Thus, radiative losses are minimal in these restricted regions (simulations are run for $\la22\tcc$).}.
We assess the consequences of restricted cooling in Appendix~\ref{appendix:extra_survival_plots}.


The simulation domain extends $100\rcl$ downwind of the cloud's initial location and $12\rcl$ along each transverse dimension.
Gas with wind properties (including magnetic fields, if $\beta$ is finite) flows into the domain $20\rcl$ upwind of the cloud's initial center of mass,
We enforce outflow conditions for the other boundaries.\footnote{ Enforcement of outflow conditions, on transverse boundaries, maintains $\nabla \cdot {\bf B} = 0$ at roughly the same precision as periodic boundaries.}

Table~\ref{tab:sims} provides a summary of our simulation properties.
We primarily employ low-resolution ($\rcl/\Delta x = 8$) simulations for our non-radiative parameter study.
The primary simulations we use to assess the cloud-wind interactions including radiative cooling have $\rcl/\Delta x = 64$.
We briefly investigate the impact of resolution in Appendices~\ref{appendix:extra_survival_plots} and \ref{appendix:convergence}.

Finally, we note that our simulations employ a reference frame tracking scheme.
Unfortunately, an implementation bug caused our $\rcl/\Delta x < 64$ simulations to effectively have no frame tracking (compared to \vw, the frame velocity is near-zero).
However, our $\rcl/\Delta x = 64$ simulations used an improved version of the code in which the frame velocity was properly updated every $0.0625\tcc$.
In Appendix~\ref{appendix:frame_tracking} we provide more details about the scheme and show that the differences have a negligible impact on our result.

\subsection{\Kdot\ Calculation}
\label{section:Kdot_method}

As a post-processing step, we estimate $\Kdotavg$ as a function of $K$ and $t$ for each of our simulations.
Recall that $\Kdot(K,t)$ is the Lagrangian entropy derivative averaged over all fluid elements originating in the cloud with entropy $K$.
$\Kdotavg$ is simply $\Kdot$ averaged over some time interval $\Delta t$.

Because mass (or in this case, passive scalar mass) is conserved, we can write an analog to the continuity equation:
\begin{equation}
  \label{eqn:Kcontinuity}
  \frac{\partial}{\partial t}\left(\dMpsdK\right) + \frac{\partial}{\partial K}\left(\Kdot \dMpsdK\right) = 0.
\end{equation}
This describes the changes in passive scalar mass profile as a function of $K$ for a pair of snapshots measured at $t^n$ and $t^{n+1}$.

Consider a set of discrete $K$ bins where the $i$th bin has center $K_i$, width $\delta K_i$, and encloses a passive scalar mass of $m_{{\rm ps},i}$.
Integrating equation~\ref{eqn:Kcontinuity} in time from $t^n$ to $t^{n+1}$ and over the $i$th $K$-bin (from $K_{i-1/2}=K_i-0.5\delta K_i$ to $K_{i+1/2}=K_i+0.5\delta K_i$) yields
\begin{equation}
  \label{eqn:Kintegral}
  \frac{m_{{\rm ps},i}^{n+1} - m_{{\rm ps},i}^n}{t^{n+1} - t^n} =
  \left( \overline{\Kdot \dMpsdK} \right)_{i-1/2}^{n+1/2} -
  \left( \overline{\Kdot \dMpsdK} \right)_{i+1/2}^{n+1/2},
\end{equation}
where $(\overline{\Kdot \dMpsdKinline})^{n+1/2}$ is time-averaged between $t^n$ and $t^{n+1}$.
By selecting a minimum bin, $K_1$, such that $\Kdot_{1/2} =0$ at all $t$, we can compute $(\overline{\Kdot \dMpsdKinline})^{n+1/2}$ at all bin interfaces from changes in the measured profiles.


Finally, if we assume that $(\dMpsdKinline)_{i+1/2}$ is near constant between $t^n$ and $t^{n+1}$ then
\begin{equation}
\label{eqn:Kdotavg}
\Kdotavg_{i-1/2}^{n+1/2}\approx \left.\left( \overline{\Kdot\dMpsdK} \right)_{i-1/2}^{n+1/2} \middle/ \left(\dMpsdK\right)_{i+1/2}^{n+1/2}\right. .
\end{equation}
We approximate $(\dMpsdKinline)_{i+1/2}^{n}$ via linear interpolation of the average $\dMpsdKinline$ from adjacent bins, and then average the values from $t^n$ and $t^{n+1}$ to estimate $(\dMpsdKinline)_{i+1/2}^{n+1/2}$.
To enforce our assumption, we focus on \Kdotavg\ measurements where $f$, given by 
\begin{equation}
    f = \frac{|(\dMpsdKinline)_{i+1/2}^{n+1}-
               (\dMpsdKinline)_{i+1/2}^{n}|}
             {\min((\dMpsdKinline)_{i+1/2}^{n+1},
                   (\dMpsdKinline)_{i+1/2}^{n})},
\end{equation}
is less than 0.25 and omit measurements altogether where $f>2$.
While our measurements of $\Kdotavg$ may be somewhat biased, the overall dependence on $K$ and $t$ is still useful, particularly when the dependence is stable in time.

In practice, we estimate \Kdotavg\ from pairs of snapshots satisfying $t^{n+1} = t^n+\tcc/2$.
If the passive scalar advects into or out of the simulation domain, then Equation~\ref{eqn:Kcontinuity} should include an additional source or sink term.
Thus, we only consider snapshots at times before $1\%$ of the initial passive scalar mass escapes the domain.
We note that vorticity near the transverse outflow boundaries can introduce artificial passive scalar inflow. 
In practice, this is only an issue for \simMachThreeQuarter\ and we conservatively discard all data from that run measured after $7 \tcc$.

Appendix~\ref{appendix:convergence} includes a brief study on how resolution affects measurements of the passive scalar mass profile and our measurements of $\Kdotavg(K,t)$.
Because the profile's evolution is most sensitive to resolution for $K$-bins holding under $1\%$ of the total passive scalar mass $M_{\rm cl,0}$, our subsequent analysis primarily focuses on 
$\Kdotavg_{i-1/2}^{n+1/2}$ measurements where $m_{i-1}^n$, $m_i^n$, $m_{i-1}^{n+1}$, and $m_{i}^{n+1}$ are all $\geq0.01M_{\rm cl,0}$.

For simulations with no cooling or inefficient cooling, the dependence of our $\Kdotavg$ measurement on $K$ and $t$ are remarkably robust with respect to resolution.
At the same time, measurements for simulations with rapid cooling are less robust.
In the appendix, we argue that this isn't surprising given our measurement method, and that the measurements are adequate for conveying the utility of our mixing model.
In these simulations, mixing would probably be better characterized by measurements of the average Lagrangian entropy derivative for all fluid elements in the system (rather than just those originating in the cloud).

\section{Non-Radiative Parameter Study Results}
\label{section:results-no-cool}

\subsection{$P-K$ Phase Space}
\label{section:P-KSpace}

%

In this section, we apply our formalism to a suite of (magneto)hydrodynamic non-radiative simulations that probe a wide variety of properties.

Figure~\ref{figure:p-K_phase_hydro} illustrates the time evolution of the passive-scalar weighted $P-K$ distribution for three hydrodynamic simulations with varying $\chi$ (\simFid, \simContrastThree, \simContrastTen).
Unlike the panels in the center-right column of Figure~\ref{figure:example_evo}, these only show phase distributions for fluid elements originating in the cloud, depict lower resolution simulations, and have rescaled entropy axes.
Throughout this work, we plot entropy as $\log_\chi(K/\Kcl)$ to remove most of its  $\chi$ dependence.
With this definition, it spans values from 0 through $\gamma=5/3$.

The figure indicates that the initial shock does not significantly alter the entropy of the cloud.
Instead, mixing is the primary source of entropy and gradually moves the fluid elements up from \Kcl\ to \Kw\ (denoted by black circles).
Fluid elements that have already exited the domain should generally have comparable or larger $K$ to the remaining ones since they mixed faster.
The higher $\chi$ simulations lose fluid elements more quickly because they have larger \vw.
The dotted black line denotes $\rhocl/3$, a density threshold commonly used to identify cloud mass \citep[e.g.][]{scannapieco15a,schneider17a, gronke18a}.
The figure shows that the vast majority of fluid elements cross this threshold by \tsim{4.5\tcc}, (as  discussed in Appendix~\ref{appendix:convergence} there is some resolution dependence), which is consistent with prior work \citep[e.g.][]{sparre18a}.
The $\log_\chi$ scaling clearly removes most of the $\chi$ dependence dependence in the entropy evolution.

Each simulation's $P$ evolution follow a common evolution, largely independent of $\chi$.
In each case, the shock initially produces large $P$ perturbations. 
By $2.5\tcc$, the motions giving rise to the under-pressured gas have been slightly damped. While not shown, the distribution's extent is stable between \tsim{1.5\tcc} and \tsim{3.5\tcc}, albeit with minor fluctuations in the minimum.
The low pressure gas at these times is presumably supported by the vorticity produced by the initial shock, the post-shock flow in the shearing layer (at the cloud boundary), and the formation of the vortex rings \citep{klein94a}.
Between $3.5\tcc$ and $4.5\tcc$ the minimum pressure drops once more and subsequently all pressure perturbations damp away. 

The $\chi$-dependence manifests in the $P$ distribution in two main ways. First, the minimum pressure at $4.5\tcc$ is larger for $\chi=100$ than it is in the other cases.
While it's unclear how robust this difference is, we note that \citet{klein94a} reported that the post-shock flow was the primary generator of vorticity in their $\chi=10$, $\Mw=0.9$, 2D ellipsoidal cloud simulation and argued that it scales with \tsim{\chi^{1/2}}.

The other difference, is that the mode of the $\chi=1000$ pressure distribution has a positive offset at $8.5\tcc$ (the mode is roughly double the initial pressure at $9.5\tcc$).
This is an unexpected artifact caused by the reflection of waves and discontinuities off of the transverse outflow boundaries.
We reran these simulations with three times larger transverse widths and found that this artifact is first noticeable at \tsim{6\tcc}.
Furthermore, while all three simulations were affected, the magnitude strongly scaled with $\chi$.
We don't expect this effect to significantly influence our results because subsequent analysis of \simContrastTwo, \simContrastThree, and \simContrastTen\ (our only $\chi>100$ simulations) ignores data at $t/\tcc>6,\, 5.5,\, {\rm and}\, 4.5$ (more than $1\%$ of the passive scalar leaves the domain by these times).

Having explored the impact of $\chi$\ on the gas distribution, we now consider the effect of magnetic fields. Figure~\ref{figure:p-K_phase_mhd} illustrates the phase evolution of the fluid elements originating in the cloud for three simulations with transverse magnetic fields of different strengths (\simBTen, \simBTHundred, \simBThousand).
In contrast to the pure hydro cases, cloud destruction proceeds far more slowly when $\beta=10$. 
It takes longer for fluid elements to cross the $\rhocl/3$ threshold and at $t=8.5 t_{cc}$, a much larger fraction of the fluid elements have $K<\Kw$ (and lie within the simulation domain). 

As $\beta$ increases, the clouds are more readily destroyed and the phase distributions bear greater resemblance to those of the pure hydro simulations.
The minimum pressure appears to correlate with the initial $\beta$ at early times.
This suggests that its supported by magnetic stress.
Because the magnetic fields are initially transverse, we expect the shock that propagates through the cloud to transfer energy to the magnetic field, thereby elevating the magnetic pressure.
We expect the pressure to be less supported by vorticity (than in the purely hydrodynamical case) because magnetic fields impede its growth.

Both figures clearly illustrate that much of the interesting evolution of the cloud-crushing
problem occurs over the $K$ dimension. While there is variation in the pressure, its both a transient effect that largely fades away at late times, and is smaller than the variation in $K$.

\begin{figure}
  \center
\includegraphics[width = \columnwidth]{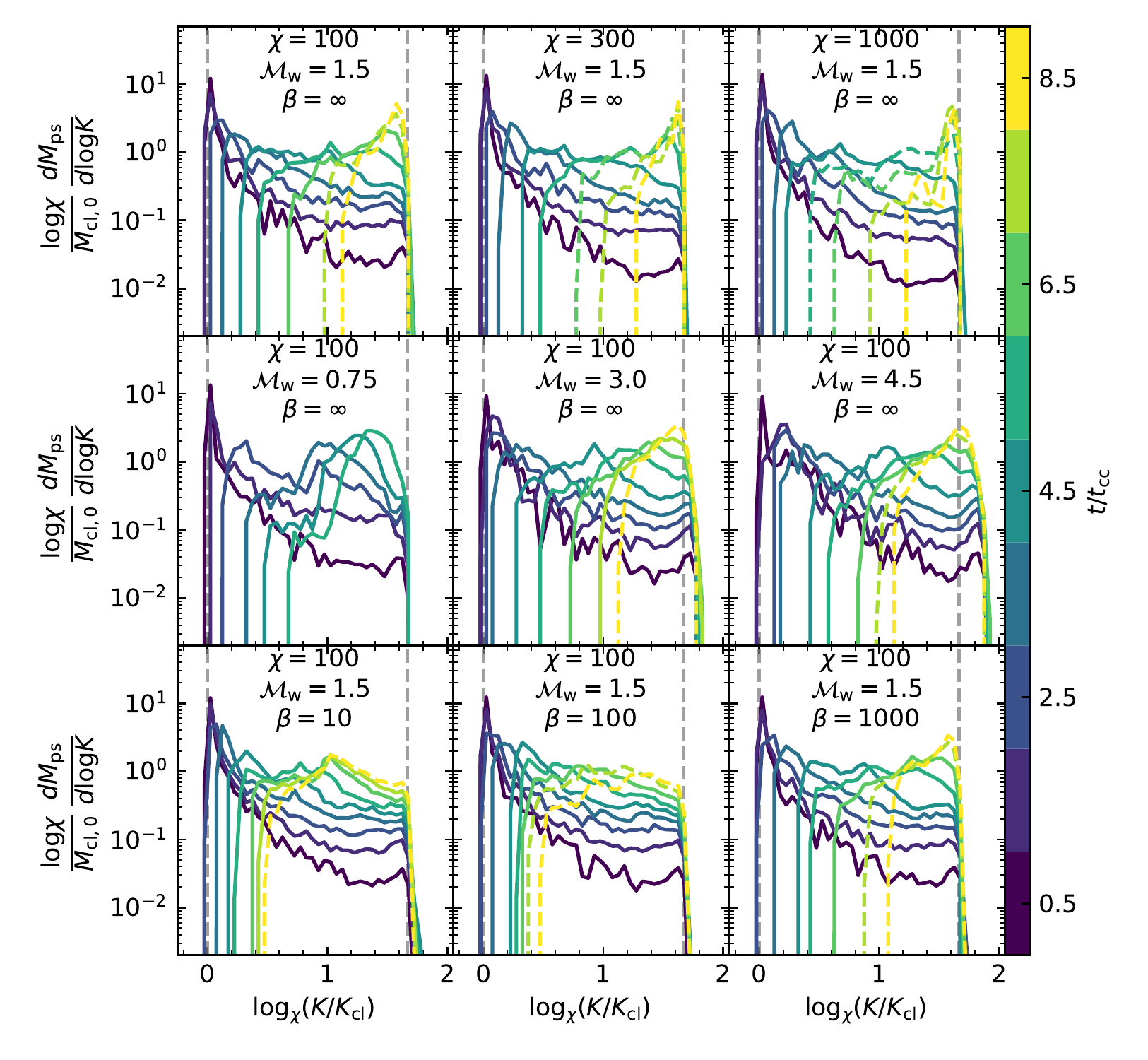}
\caption{\label{figure:dMpsdK} 
  Temporal evolution of the mass profile as a function of entropy for fluid elements originating in the cloud, $d\Mps/d \log_\chi K$.
  The profiles are normalized by the initial cloud mass.
  Each panel depicts a different simulation;
  $\chi$, $\Mw$, and $\beta$ vary across the top, middle, and bottom rows.
  Colors denote measurement times and dash lines denote profiles measured after $1\%$ of the passive scalar leaves the domain.
  Gray vertical dashed lines denote \Kcl\ and \Kw (the initial entropy of fluid elements in the cloud and wind).
  The scaling of time (in terms of \tcc) and entropy largely removes the effects of $\chi$ and \Mw\ (where $\Mw\geq1.5$) for pure hydrodynamic mixing.
  However, it does not account for differences in subsonic winds.
  At late times, the distributions' extents convey that increasing magnetic field strengths more effectively impede mixing.
}
\end{figure}

%

\subsection{Passive Scalar Mass Distribution over $K$}
\label{section:KSpaceDist}

Having qualitatively established that the bulk motion of cloud fluid elements through $P-K$ space both occurs primarily along $K$ and reflects the initial conditions and modelled physics, we now consider a more quantitative parameterization of mixing. 

If we assume that pressure perturbations are broadly unimportant for the system's evolution, we can integrate over the phase distribution's pressure dependence to get $\dMpsdKinline$.
We effectively trade the information encoded in the pressure perturbations for a dimensionality reduction.
Recall that $(\dMpsdKinline)dK$ specifies the mass of all fluid elements with entropy between $K$ and $K+dK$.
Figure~\ref{figure:dMpsdK} illustrates the time evolution of $(\dMpsdKinline)dK$ for a selection of times for each non-radiative simulation listed in Table~\ref{tab:sims} with a resolution of $\rcl/\Delta x=8$.

At $t=0.5\tcc$, each simulation's profile has a peak at \Kcl\ and a long tail extending to \Kw.
Over time, mixing increases the entropy of the fluid elements near the lower edge of the histogram, $K_{\rm min}$. 
By $t=3.5\tcc$, $K_{\rm min}$ starts to increase, indicating that all of the fluid elements from the cloud have started mixing.
We largely ignore differences in the profiles at intermediate and late times that are depicted by dashed lines because different fractions of passive scalar remain in the simulation domain when those are measured.

For the hydrodynamical simulations (top two rows of Figure~\ref{figure:dMpsdK}), the narrow histograms near \Kw\ at $t=8.5\tcc$ reflect how the initial cloud fluid elements have largely homogenized with the wind.
The figure shows that the scaling of our $K$-bins and $t$ in terms of $\log_\chi (K/\Kcl)$ and \tcc\ almost entirely captures the evolution's $\chi$ dependence.
This $t$ scaling also largely removes the \Mw\ dependence for supersonic simulations; however, the histograms' upper edges do scale weakly with \Mw.

The subsonic run, \simMachThreeQuarter, has the most unique evolution.
In this case, mixing appears to more rapidly increase entropy below $\log_\chi (K/\Kcl)\sim0.7$ and the intermediate distributions develop a more prominent central peak.
We defer further examination of the subsonic cloud-wind interaction to future work.



Finally, we turn to the MHD simulations (bottom row of Figure~\ref{figure:dMpsdK}). 
At early times ($t\la2.5\tcc$) the profile evolution is largely the same as before, but by  $t=3.5\tcc$ the suppression of mixing by the magnetic fields causes the evolutionary paths to diverge. 
Since stronger fields (in a given configuration) more strongly suppress mixing, the rate at which $K_{\rm min}$ increases scales with increasing $\beta$.

\begin{figure}
  \center
\includegraphics[width = \columnwidth]{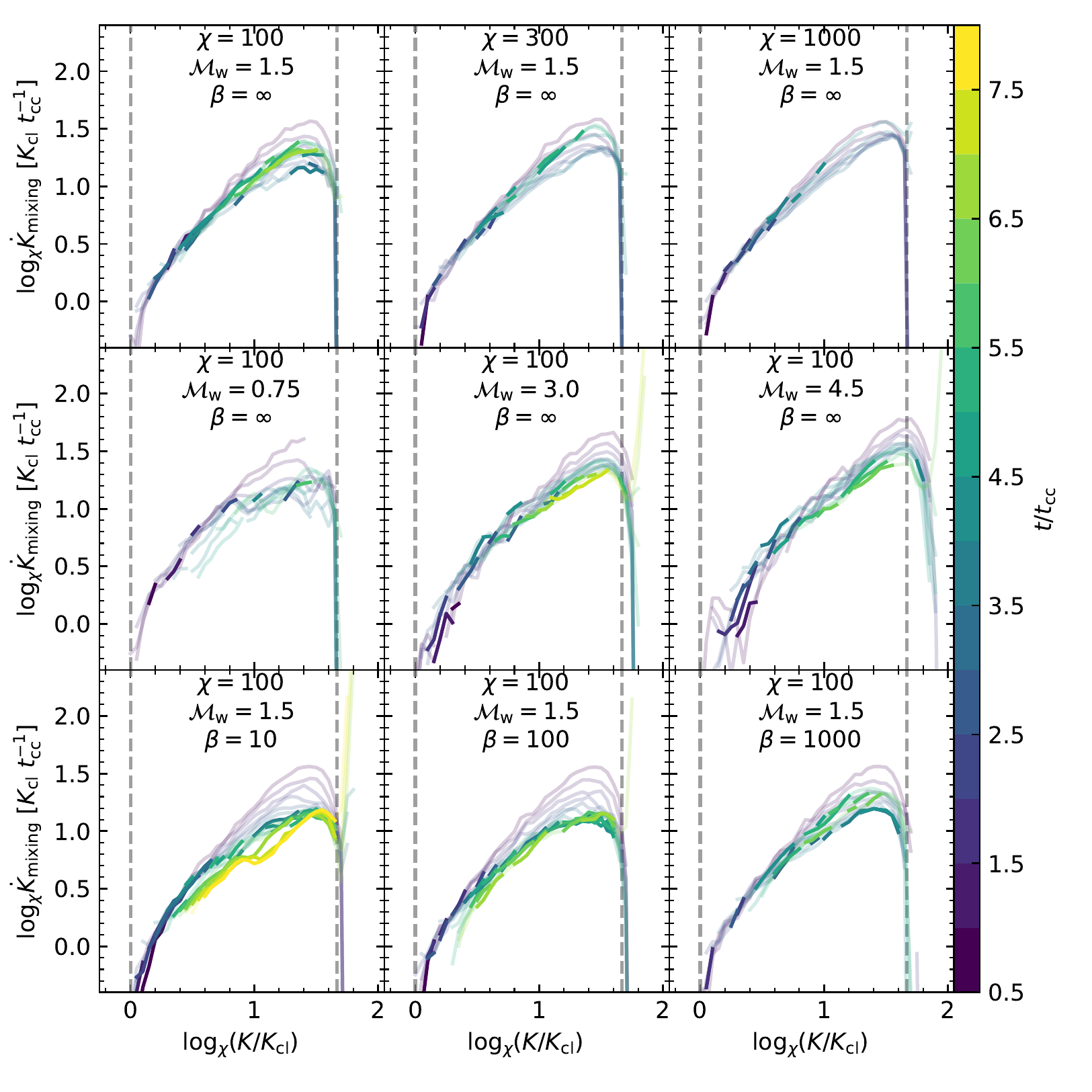}
\caption{\label{figure:KvsKdot} Passive scalar mass weighted and time averaged
  values of $\Kdotmixing(K,t)$ computed from the histograms in Figure~\ref{figure:dMpsdK}. 
  Each \Kdot\ curve is averaged over $0.5\tcc$. 
  Measurements have been omitted (made transparent) at any bin edges where the interpolated $\dMpsdKinline$ changes by more than a factor of 3 (1.25) over this interval.
  Measurements are also made transparent when they are adjacent to histogram bins containing under $1\%$ of the initial mass. 
  The curves are only computed at times when more that $99\%$ of the fluid elements originating in the cloud lies in the simulation domain.
}
\end{figure}

\begin{figure}
  \center
\includegraphics[width = \columnwidth]{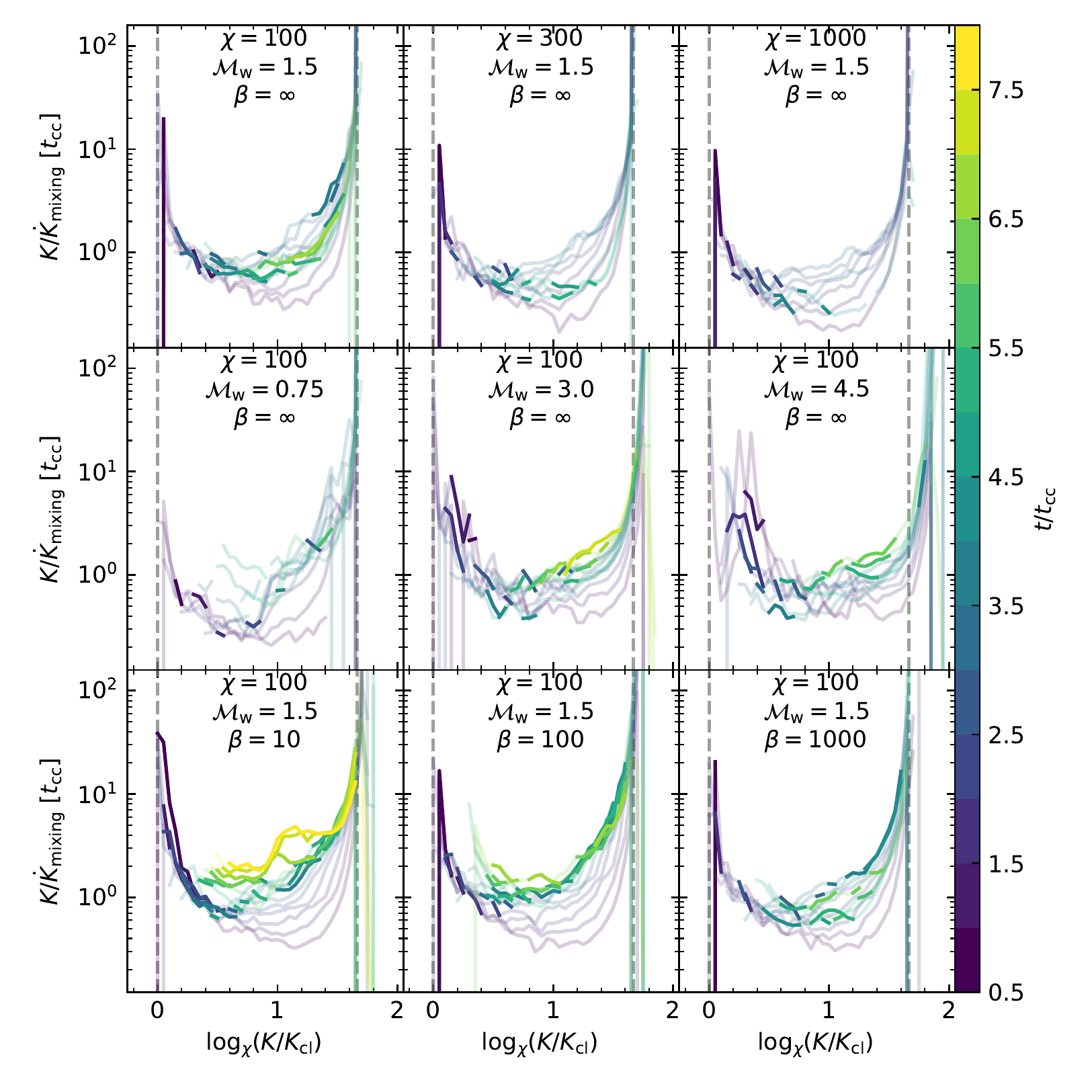}
\caption{\label{figure:mixing_t} Time scales, for mixing to double the $K$ of a fluid element that originated in the cloud. For the hydrodynamical simulations, the time scales are relatively consistent across $\chi$ times and they are of the same order as \tcc . Although the timescales for the MHD simulations are similar at early times, they get longer at late times.
The data selection is the same as in Figure~\ref{figure:KvsKdot}.
}
\end{figure}

\subsection{Mixing Rate Estimation}
\label{section:mixing_rate}

In this section, we use this distribution evolution to estimate the rate at which fluid elements from the cloud mix.
Figure~\ref{figure:KvsKdot} illustrates our measurements of $\Kdotavg(K,t)$, averaged over $\Delta t = \tcc /2$,  as functions of $K$ for each of our simulations.
See \S\ref{section:Kdot_method} for explanations of the calculation and how we identify the best measurements.
As discussed in \S\ref{section:mixing_model}, we refer to these measurements as $\Kdotavgmixing(K,t)$ because mixing is the dominant entropy generation mechanism.

In each hydro simulation, $\Kdotavgmixing/(\Kcl\tcc^{-1})$ broadly has a power law relationship in terms of $K/\Kcl$, with a near-unity slope for $0.3\la \log_{\chi} K/K_{\rm cl} \la 1.4$.
The section below $\log_{\chi} K/K_{\rm cl} \sim 0.8$ may be slightly steeper (${\sim}1.1$ for \simFid) while the upper section's slope may be slightly shallower (${\sim}0.9$ for \simFid) and could be time dependent.
Note that the top row of Figure~\ref{figure:converge_Kdot} from Appendix~\ref{appendix:convergence} suggests that these trends are robust to resolution effects.

The higher $\chi$ simulations appear to have slightly steeper slopes than \simFid, but the dearth of high quality measurements make this comparison tenuous, especially at high $K$.
The higher \Mw\ simulations have greater variance in their measurements than \simFid\ (possibly due to their stronger initial shocks), but are otherwise broadly consistent.
Without better measurements, we're unable to make any comparisons with \simMachThreeQuarter.

Next, we consider the MHD simulations. 
As in Figure~\ref{figure:dMpsdK}, the $\beta\leq100$ $\Kdotavgmixing(K,t)$ measurements only start diverging from the \simFid\ measurements at $t\sim3.25 \tcc$. 
The suppression of mixing gives $\Kdotavgmixing(K)$ a shallower slope.
As the initial field strength decreases, the measurements more closely resemble those from \simFid.


Figure~\ref{figure:mixing_t} illustrates $K/\Kdotavgmixing(K,t)$, which is the time that mixing takes to double $K$, in units of \tcc.
It makes the slope variations in \Kdotavgmixing, above and below $\log_{\chi} K/K_{\rm cl} \sim 0.8$, more apparent.
Additionally, $K/\Kdotavgmixing(K,t)$ is generally within a factor of ${\sim}2.5$ of \tcc\ in each hydro simulation.
This implies that $\Kdotavgmixing(K,t)$ and $\tcc^{-1}$ share similar $\Mw$ and $\chi$ dependence.

The results in the section broadly indicate that $\Kdotavgmixing(K,t)$ robustly characterizes cloud destruction through mixing.
For idealized conditions, our results further suggest that $\Kdotavgmixing$ doesn't have a strong $t$ dependence and we can approximate $\Kdotavgmixing(K,t)$ with a time-independent function, $\Kdotmixingchar(K)$.
In the presence of additional physical effects (e.g. the presence of magnetic fields), $\Kdotavgmixing(K,t)$ shows stronger time dependence, and improves on the description of cloud destruction offered by \tcc.
This is conveyed in Figures~\ref{figure:KvsKdot} and \ref{figure:mixing_t} for \simBTen; $\Kdotavgmixing(K,t)$ clearly captures the decreasing destruction rate, presumably caused by the tangling of magnetic fields.


\section{Results with Cooling}
\label{section:results-cool}

\begin{figure*}
  \center
\includegraphics[width = 6.in]{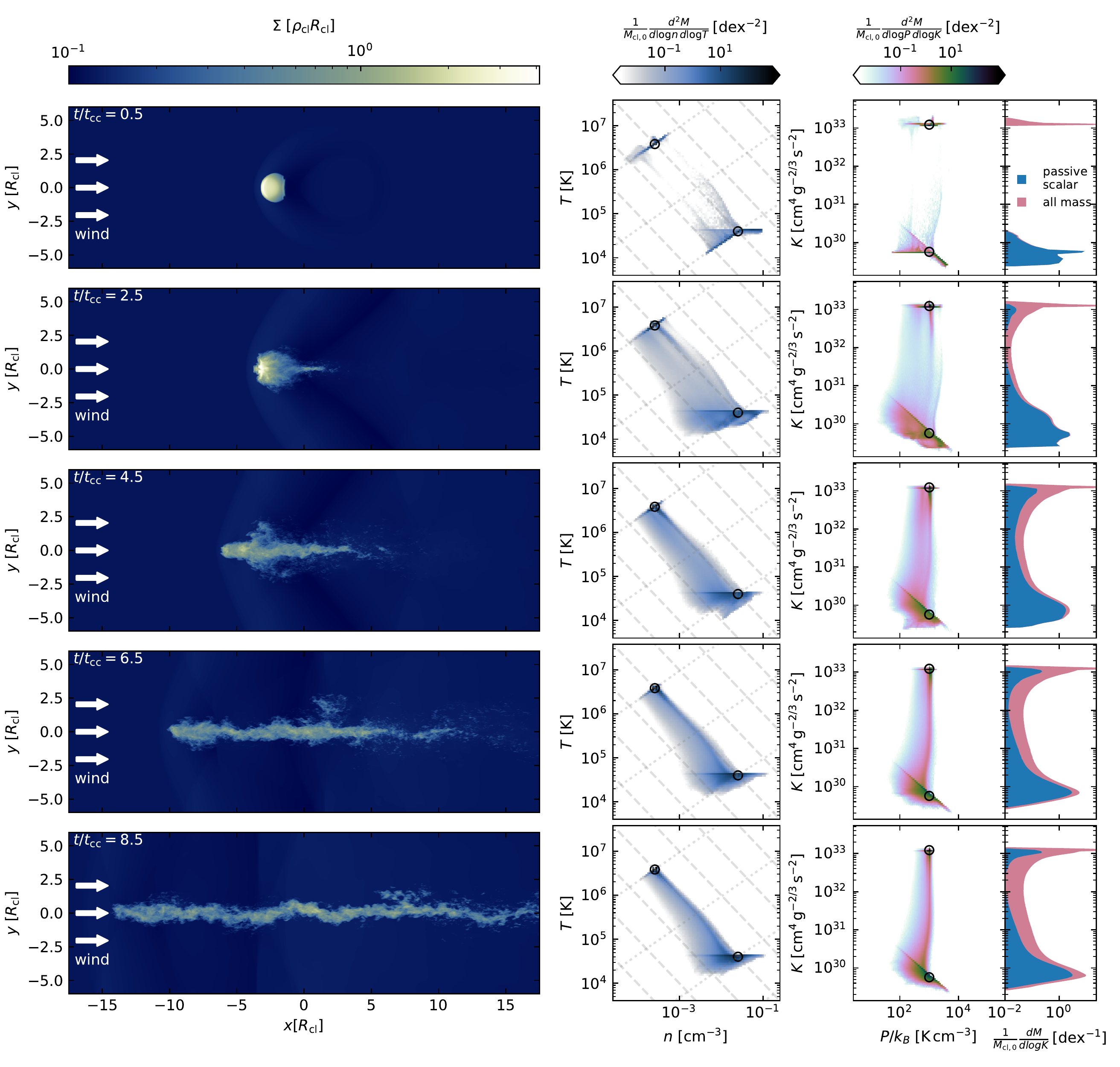}
\caption{\label{figure:example_evoRC} The same as Figure~\ref{figure:example_evo} except this shows \simRCFastFour, which shows significant cloud growth. 
  In this simulation, $\mu$ is a function of $n_H$ and $T$. 
  As in Figure~\ref{figure:example_evo}, mixing initially drives material from the colder phase to the hotter phase, but rapid cooling slows the transfer rate.
  Shortly before $4.5\tcc$, cooling causes this transfer to reverse: the cool phase starts to accrete mass.
  The stable phase diagrams are a manifestation of this growth because there's a limitless supply of hot phase gas.
  The growth is even more obvious in the right-column; by 8.5\tcc, the mass of the gas with $K<3\times10^{31}\, {\rm cm}^4\, {\rm g}^{-2/3}\, {\rm s}^{-2}$ has more than doubled.
}
\end{figure*}

\begin{figure*}
  \center
\includegraphics[width = \textwidth]{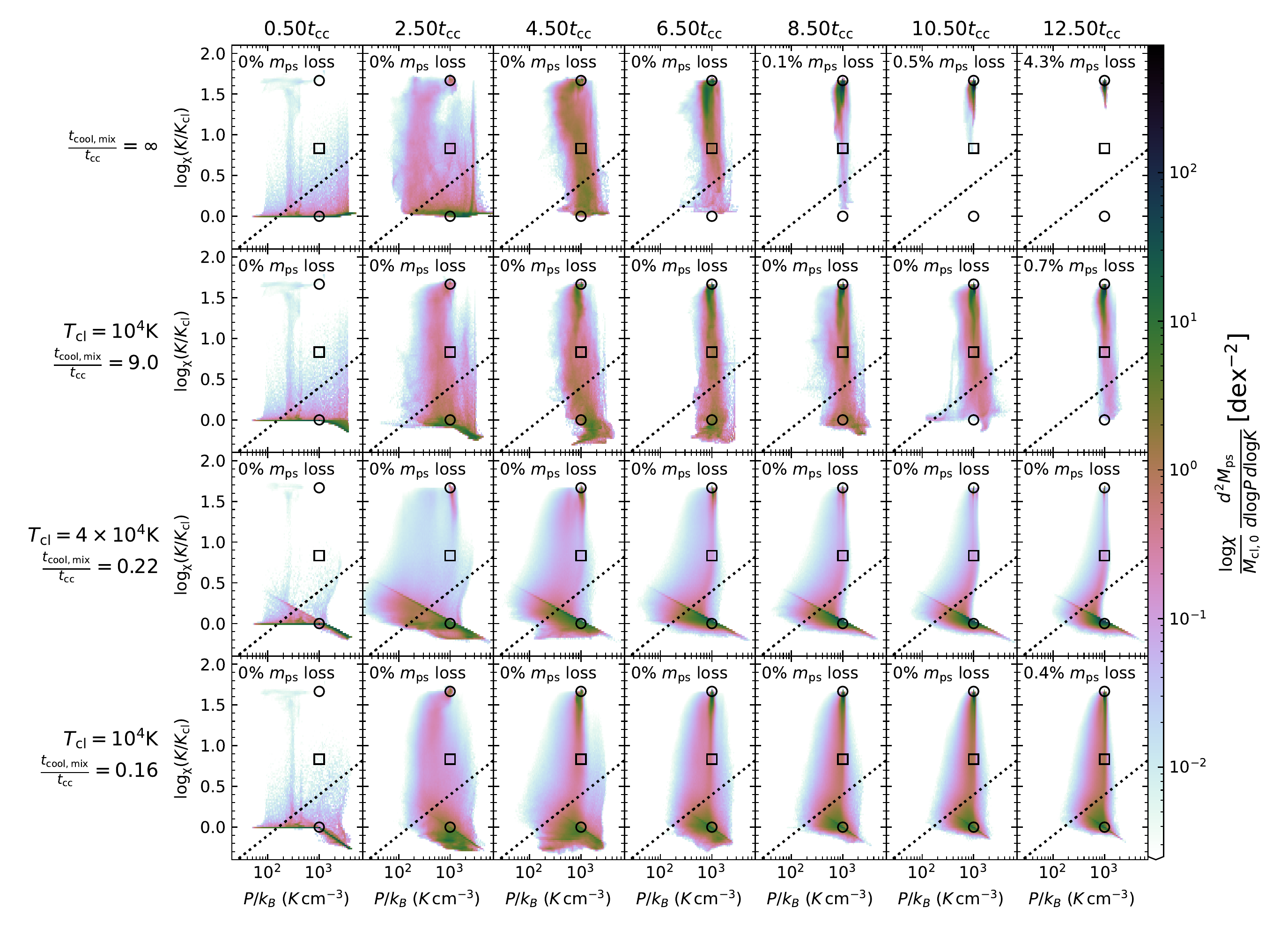}
\caption{\label{figure:pK_cooling}
  Mass weighted pressure-entropy evolution for fluid elements originating in the cloud.
  This is similar to Figure~\ref{figure:p-K_phase_hydro} except that each simulation has $\chi=100$ and $R_{\rm cl}/\Delta x = 64$.
  Instead, the radiative cooling effectiveness differs between rows. The top row has no cooling, the second row has slow cooling, and the bottom rows have fast cooling.
  At $t\geq2.5\tcc$, these distributions are qualitatively similar to the phase distributions that include all gas in the domain at low and intermediate $K$.
}
\end{figure*}

\begin{figure}
  \center
\includegraphics[width = \columnwidth]{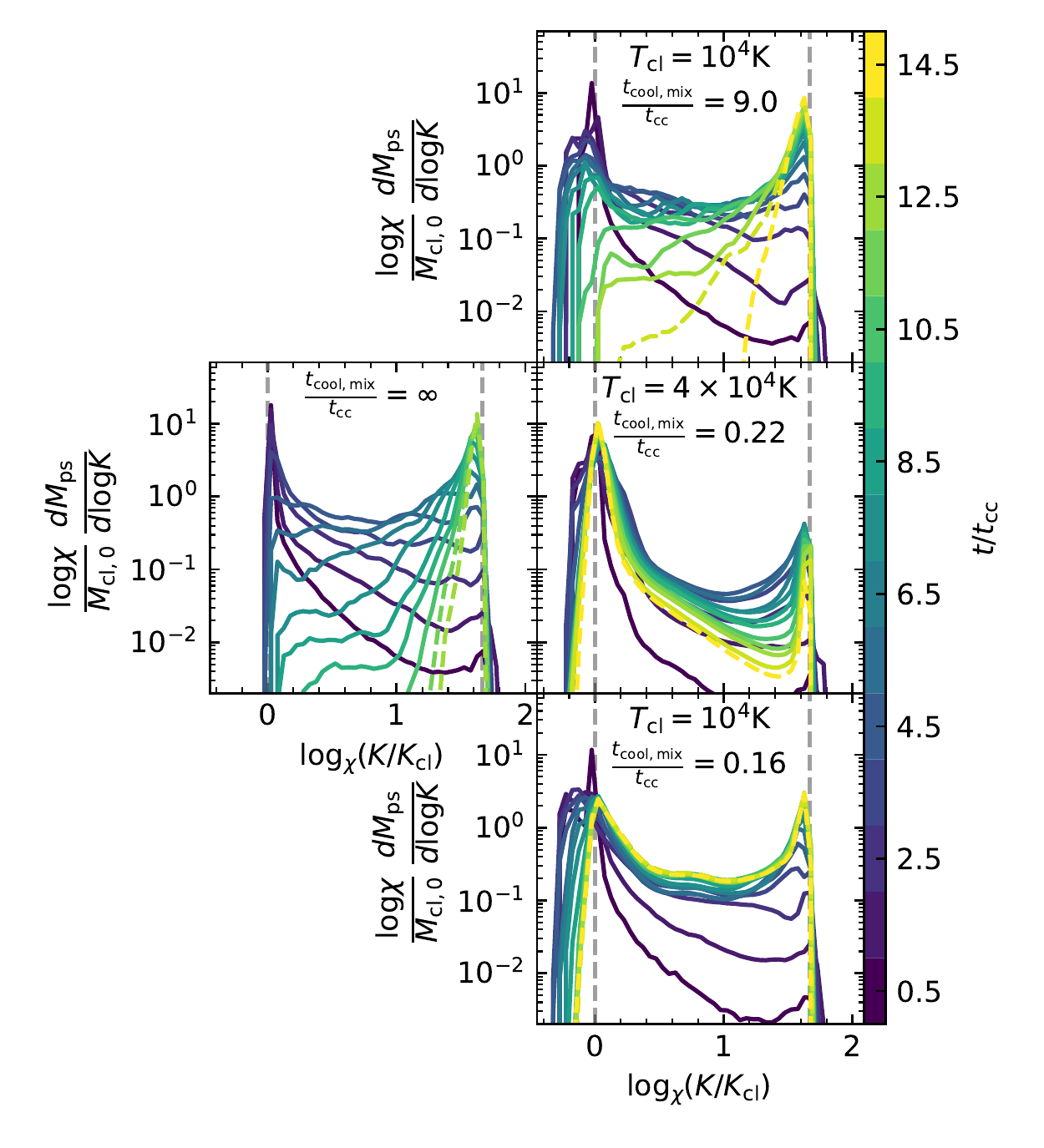}
\caption{\label{figure:dMpsdK_cool_adiabat} The one-dimensional entropy distribution as in Figure~\ref{figure:dMpsdK} except that some of the simulations include radiative cooling, and they all have $\chi=100$, $\Mw=1.5$, and $\rcl/\Delta x = 64$. The simulations are in the same order as for Figure~\ref{figure:pK_cooling}.
}
\end{figure}

\subsection{$P-K$ Phase Evolution}
\label{section:cooling_phase_evo}

Next, we apply our mixing model to hydrodynamic simulations with radiative cooling.
We consider two main regimes of cooling: slow, $\tcoolmix\sim 10 \tcc$, and fast, $\tcoolmix\sim 0.2 \tcc$.
Per \citet{gronke18a}, the cloud should be destroyed in the former case and survive in the latter.
For the fast cooling regime, we consider two separate initial cloud temperatures:  $\Tcl=10^4\,{\rm K}$ (\simRCFastOne) and $\Tcl=4\times 10^4\,{\rm K}$ (\simRCFastFour). Figure~\ref{figure:example_evoRC} depicts the latter case.
However, we only present one slow cooling simulation with $\Tcl=10^4\, {\rm K}$ (\simRCSlowOne) because \Tcl\ has minimal impact in this regime. 
We compare these simulations against the non-radiative simulation \simFid.


Figure~\ref{figure:pK_cooling} depicts how radiative cooling modifies the $P-K$ phase space distribution for fluid elements originating in the cloud.
The key takeaway is that cooling slows the spread of cloud material into the background high entropy phase.
This suppression is stronger for higher cooling rates, which reflects the fact that intermediate entropy material cools to low entropy prior to mixing with high entropy material.

Unsurprisingly, the evolution of our slow cooling case is minimally changed from the non-radiative case; the rate at which gas migrates to the high entropy phase is slower.
Rapid cooling more significantly modifies the distribution.
Consistent with our expectation of entrainment, a reservoir of gas is always present at ($p_0$,\Kcl) throughout the system's evolution.
Interestingly, after an early transient phase, which has a large scatter in $p$, the distribution approaches a near steady state.
In this state, the conditional pressure distributions have reduced scatter and a mode that that lies mostly along the $p_0$ isobar, but has a decrement near $\log_{\chi}(K/\Kcl)\sim 0.2$.

This decrement is likely an artifact of under-resolved cooling \citep{fielding20a, tan20a}.
This is supported by the fact that the decrement is almost non-existent in \simRCFastOne, where \lcool\ (see Appendix~\ref{appendix:convergence}) is actually resolved.

The over-dense diagonal line, in Figure~\ref{figure:pK_cooling}, intersecting $(p_0,\Kcl)$ lies along the isotherm corresponding to the cooling curve's temperature floor.
In the absence of this floor, the gas would cool to lower $K$. 
This is shown in Appendix~\ref{appendix:extra_survival_plots}.

Figure~\ref{figure:dMpsdK_cool_adiabat} illustrates the bulk motion of the fluid elements originating in the cloud along $K$.
For both fast cooling simulations, it shows a bi-stable medium with long-lived cold and hot gas.
During the early stages of the interaction, the colder phase loses mass to the hotter phases, but after some time this reverses.
While the exact timescale depends on \Tcl, cooling gradually becomes more effective at opposing cloud destruction.
Eventually, it is effective enough that it not only prevents loss of additional mass but also recaptures the lost mass.
This behavior manifests over a notably shorter timescale for \simRCFastFour.

\subsection{Turbulent Radiative Mixing Layer Entrainment}
\label{section:entrainment}

\begin{figure}
  \center
\includegraphics[width = \columnwidth]{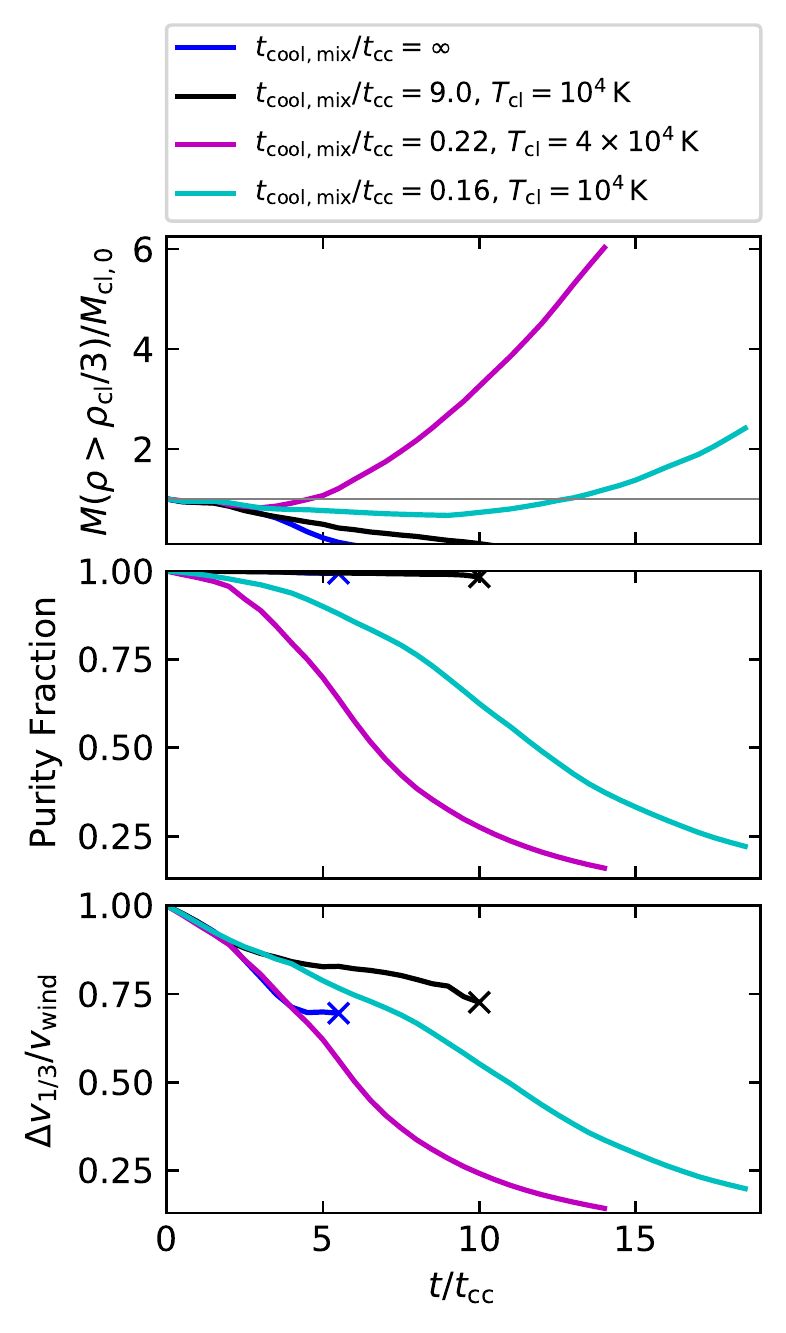}
\caption{\label{figure:bulk_evo} Evolution of total mass (top), purity fraction (middle; fraction of mass originating in the cloud), and average velocity (bottom; $\Delta v=\vw - v_{\rm cl}$) of cells satisfying $\rho>\rhocl/3$ for the simulations shown in Figure~\ref{figure:pK_cooling}.
  The ``x'' markers in the bottom two panels indicate when a cloud's mass drops to $10\%$ of its initial mass.
  During the destruction of the non-radiative and slow cooling cases (rapid mass loss), the purity fraction remains near unity, which signals that ram pressure drives acceleration.
  The accretion of mass in the rapid cooling cases causes the purity fraction to drop.
  The correlation of purity fraction and $\Delta v/\vw$ indicate that mixing drives acceleration.
}
\end{figure}

\begin{figure}
  \center
\includegraphics[width = \columnwidth]{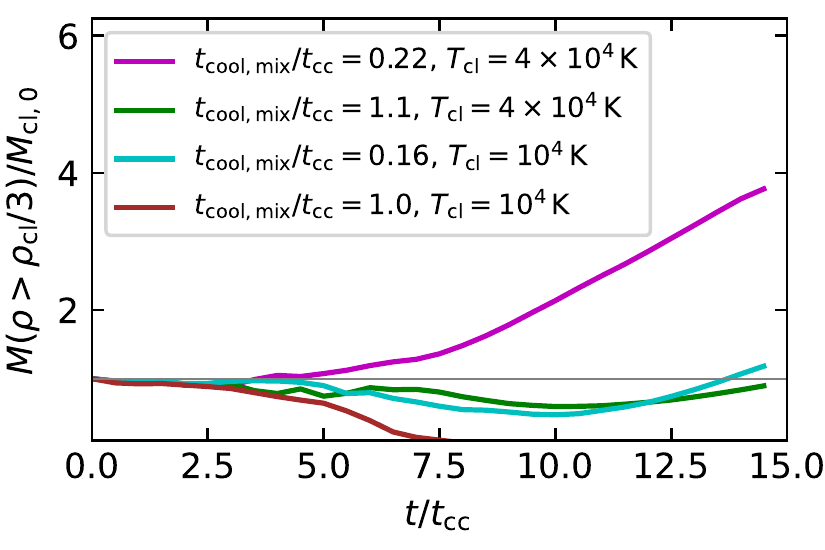}
\caption{\label{figure:extra_survival}
Dependence of cold phase mass growth on \Tcl.
To make the comparison as fair as possible, all data was measured from simulations with $\chi=100$, $\Mw=1.5$, $p/k_B = 10^3\, {\rm cm}^{-3}\, {\rm K}$, and $\rcl/\Delta x=16$. 
The magenta and cyan curves are measured from lower resolution versions of \simRCFastFour\ and \simRCFastOne, while the green and brown curves have radii of 1000 pc and 9 pc.
Note, we have verified that the brown curve goes to zero (the y-axis starts at 0.1).
This demonstrates that the $\tcoolmix/\tcc$ criterion alone doesn't fully specify the cloud evolution.
}
\end{figure}

\begin{figure}
  \center
\includegraphics[width = \columnwidth]{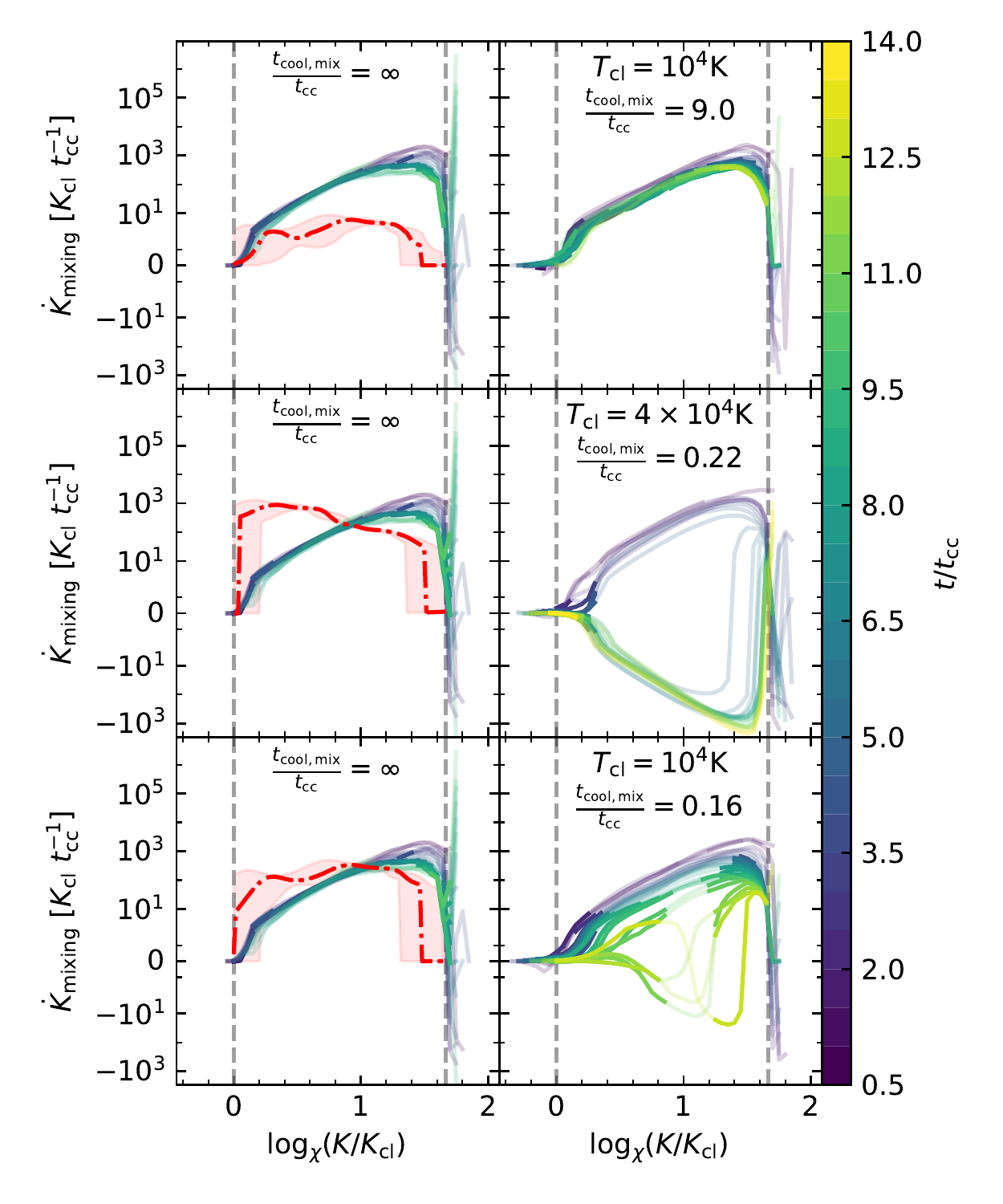}
\caption{\label{figure:dKdt_cool_adiabat}
  Comparison of passive scalar mass weighted and time averaged $\Kdot(K)$ measured with and without radiative cooling for three sets of physical conditions with $\chi=100$, $\Mw=1.5$ and $\rcl/\Delta x = 64$.
  The {\bf left} panels compare measurements for just adiabatic mixing ($\Kdotmixing(K)$) against the predicted contributions from cooling ($\Kdotcool(K)$), while the {\bf right} panels depict measurements involving both mixing and cooling.
  The red dashed line shows $-\Kdotcool(K)$ for the initial pressure and the shaded regions show variations from 0.5 dex pressure perturbations.
  The other curves are computed from the histograms in Figure~\ref{figure:dMpsdK_cool_adiabat} (only one set of $\Kdotmixing(K)$ curves are shown).
  Curves are omitted when at least $1\%$ of the fluid elements originating in the cloud have left the domain. Curve segments adjacent to bins with under $1\%$ of the initial cloud mass are transparent.
}
\end{figure}

Figure~\ref{figure:bulk_evo} shows how the different cooling regimes affect the bulk property evolution of the colder, denser phase (gas with $\rho>\rhocl/3$).
The top panel illustrates the total mass evolution and confirms that the \citet{gronke18a} criterion accurately predicts the cloud's fate.
The cloud is destroyed in both the non-radiative and slow cooling cases, although cooling slows the destruction rate.
On the other hand (as noted in \S\ref{section:cooling_phase_evo}), in the fast cooling cases, the cloud not only survives but also starts to rapidly grow in mass.

The bottom and middle panels depict the evolution of the velocity and purity fraction (i.e. the cold phase mass fraction of fluid elements initialized in the cloud).
The correlation in the evolution of velocity and purity fraction reflects an inelastic collision in the fast cooling limit; this is expected for turbulent radiative mixing layer entrainment \citep[][Tonnesen \& Bryan, in prep.]{gronke18a,schneider20a}.
The sustained high purity fraction signals that a different process, probably ram pressure, dominates acceleration in the non-radiative and weak cooling regime.
Note that the minor offset in the velocity and purity fraction evolution suggests that ram pressure could play a subdominant role in the fast cooling limit.

Interestingly, Figure~\ref{figure:bulk_evo} also indicates that the $\tcoolmix/\tcc$ criterion alone doesn't fully specify the cloud evolution.
In the fast cooling limit, the rate of cloud growth depends on \Tcl ; \simRCFastOne\ takes at least twice as long as \simRCFastFour\ to show growth despite having nearly identical $\tcoolmix/\tcc$ ratios.
This depressed cloud growth in \simRCFastOne\ is accompanied by a delay in the time at which the cloud is entrained.

Figure~\ref{figure:extra_survival} underscores the significance of this difference in growth.
 The green and brown curves show the mass growth of simulations that are respectively identical to \simRCFastFour\ and \simRCFastOne, except that they have $\tcoolmix/\tcc\sim1$.\footnote{This difference in \tcc\ is achieved by reducing \rcl\ by a factor of five.}
The green curve shows nearly identical growth to \simRCFastOne\ (shown in cyan), despite the difference in $\tcoolmix/\tcc$.
As we'll conclude below, this difference in growth arises from differences in $\tcool/\tcc$ between \Tcl\ and \Tmix\ (for reference the minimum $\tcool/\tcc$ for the green curve is half of that for \simRCFastOne).
Moreover, the fact that the brown curve goes to zero, despite having a comparable $\tcoolmix/\tcc$ to the green curve, illustrates that this difference can even modify the survival cloud survival criterion.

To interpret these results, we consider them in terms of our mixing model.
For each cooling case, Figure~\ref{figure:dKdt_cool_adiabat} compares standalone non-radiative $\Kdotavgmixing(K,t)$ measurements and the $\Kdotcool(K)$ prediction against the $\Kdotavgtotal(K,t)$ measurements from the simulations with radiative cooling.

Given our result from \S\ref{section:mixing_rate} that $K/\Kdotmixing(\Kmix,t)\sim\tcc$ for most times when $(\dMpsdKinline)(\Kmix)>0$, the survival criterion $\tcc>\tcoolmix$ can be directly visualized in terms of this plot.
Turbulent radiative mixing layer entrainment is expected when $-\Kdotcool$ exceeds the characteristic value of $\Kdotmixing$ at $\Kmix=\chi^{5/6}\Kcl$.
Given $\Kdotmixingchar(K)\sim K/\tcc$ and $|\Kdotcool|$'s mostly inverse dependence on $K$ for realistic ISM conditions, satisfaction of the survival criterion basically guarantees that $\Kdotmixingchar (K) < \Kdotcool(p_0,K)$ for $K\in[\Kcl,\Kmix]$.
Thus, we predict a negative $\Kdottotal(K,t)$ over that interval, and, by extension,  entrainment.

The outcome of the slow cooling case ($\tcoolmix/\tcc \sim 10$) is clear-cut.
Because $|\Kdotmixingchar(K)|\gg |\Kdotcool(p_0, K)|$, $\Kdottotal(K,t)$ resembles $\Kdotmixing(K,t)$ and the cloud is destroyed.
Likewise, the ultimate fates in the fast cooling cases are also predictable.
However, the detailed shape and temporal evolution of $\Kdottotal(K)$ is less straightforward. 

The $\Kdottotal(K,t)$ evolution for the cool cases directly reflects the discussion from the previous section (\S\ref{section:cooling_phase_evo}).
In the earliest stages of the interaction, $\Kdottotal(K,t)$ is positive because radiative cooling is unable to prevent initial mixing of the phases.
As the process continues, $\Kdottotal(K,t)$ gradually decreases with time, which indicates that cooling becomes more effective at combating mixing.
Eventually, the opposition of cooling to mixing becomes so effective that it reverses the transfer of gas between phases, which causes $\Kdottotal(K,t)$ to become negative.

It is around this time that our measurements of $\Kdottotal(K,t)$ lose meaning, since an increasing fraction of the colder phase is composed of gas originating in the hot phase.
This is discussed in further in Appendix ~\ref{appendix:convergence}.
Nevertheless, we have included the measurements because they illustrate, if imprecisely, the expected behavior.

We now consider why \simRCFastFour\ begins rapid growth in roughly half the time as \simRCFastOne.
Figure~\ref{figure:dKdt_cool_adiabat} suggests that this difference derives from the local shape of the cooling curve.
The major difference is that $|\Kdotmixing(K)/\Kdotcool(K)|$ is roughly an order of magnitude smaller in $0.1\la \log_\chi (K/\Kcl) \la 0.6$ for \simRCFastFour.
Phrased another way, the value of \tcool\ near the entropy (or temperature) of the colder phase appears to determine how rapidly the cloud grows.

\subsection{Cooling Curve Variations}
\label{section:custom_curves}

\begin{figure}
  \center
\includegraphics[width = \columnwidth]{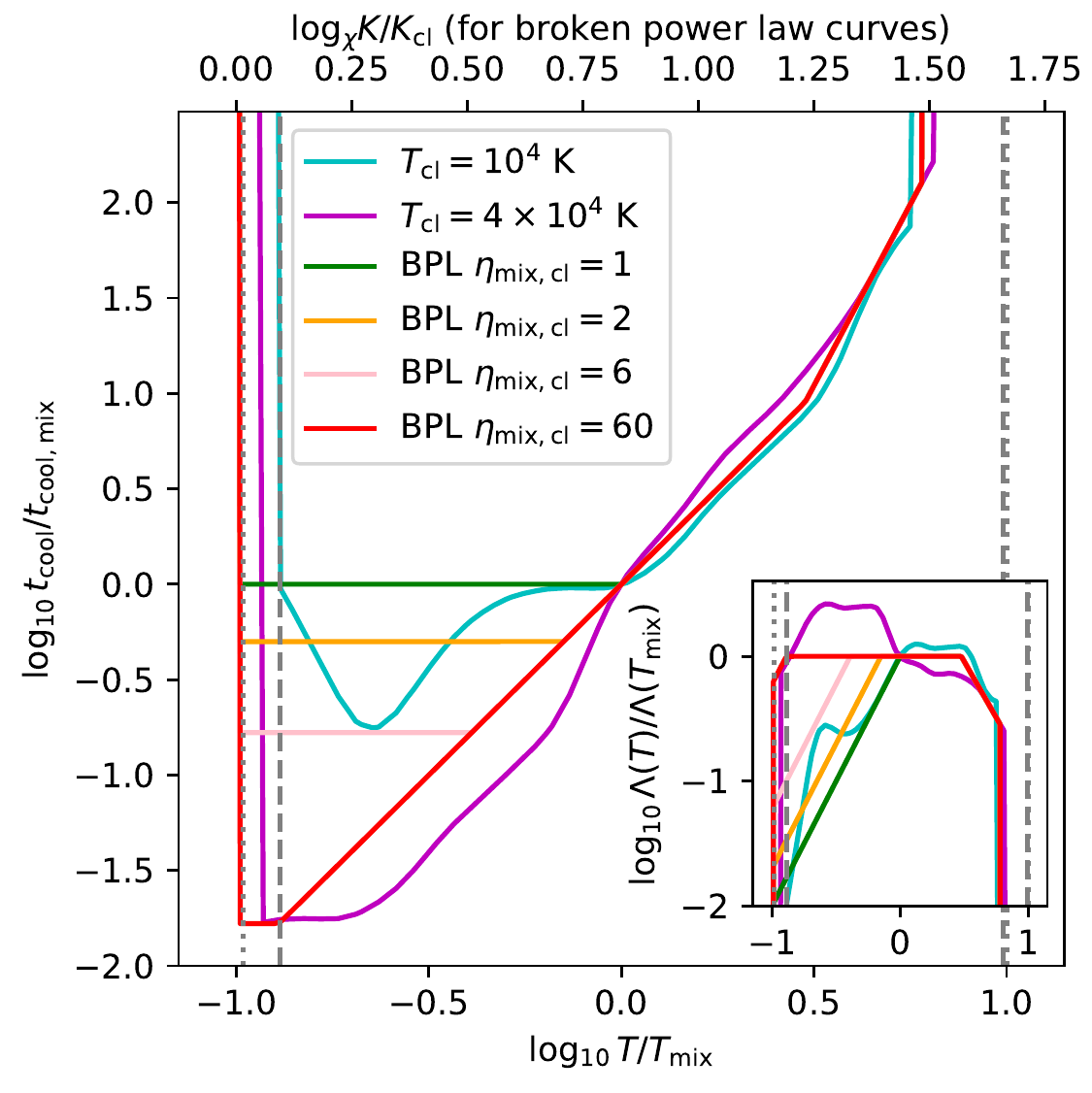}
\caption{\label{figure:custom_curves}
  Comparison of the truncated cooling curves, measured at $p/k_B=10^3\, {\rm K}\, {\rm cm}^{-3}$ and normalized by their properties at \Tmix, used in our ($\chi=100$) radiative cooling simulations.
  The upper axis denotes the entropy for just the broken power-law cooling curves, which assume a fixed $\mu$ of 0.6.
  In all displayed curves, \Kcl\ ($=0.1^{\gamma}\Kmix$), $\Kmix$, and \Kw\ ($=10^{\gamma}\Kmix$) always coincide with \Tcl, \Tmix, and \Tw, by definition.
  The gray dashed (dotted) lines denote the relative locations of \Tcl\ and \Tw\ when $\Tcl = 10^4\, {\rm K}$ ($4\times 10^4\, {\rm K}$).
  For each broken power law cooling function, $\Tcl =0.1\Tmix$ and $\Tw =10\Tmix$.
}
\end{figure}

\begin{figure}
  \center
\includegraphics[width = \columnwidth]{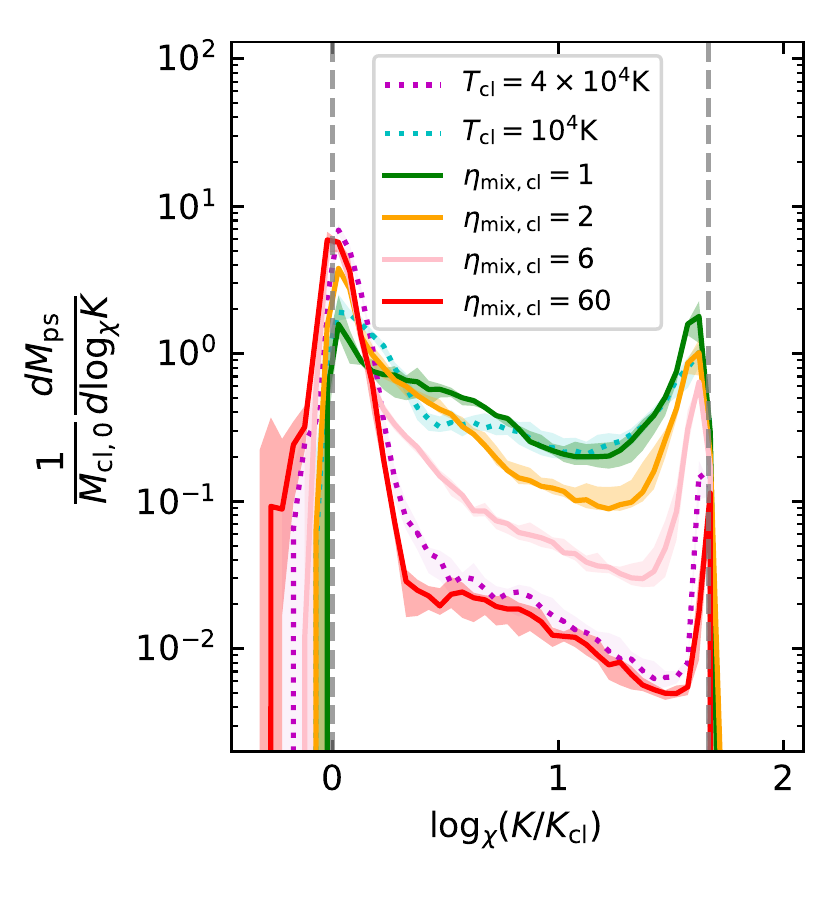}
\caption{\label{figure:dMpsdK_custom_curves} 
Late time entropy distributions of fluid elements originating in the cloud for simulations run with the custom broken power-law cooling functions (see Figure~\ref{figure:custom_curves}). 
The shaded region depicts entire temporal variation during $t/\tcc\in [10,13.5]$ and the colored lines denote the median.
For comparison, distributions are also shown for $\rcl/\Delta x=8$ versions of \simRCFastFour\ and \simRCFastOne.
At least $99\%$ of the passive scalar remain in the domain through $13.5\tcc$ in nearly all cases.
However, this is only true for \simBPLawOne, through $11\tcc$. 
}
\end{figure}

\begin{figure}
  \center
\includegraphics[width = \columnwidth]{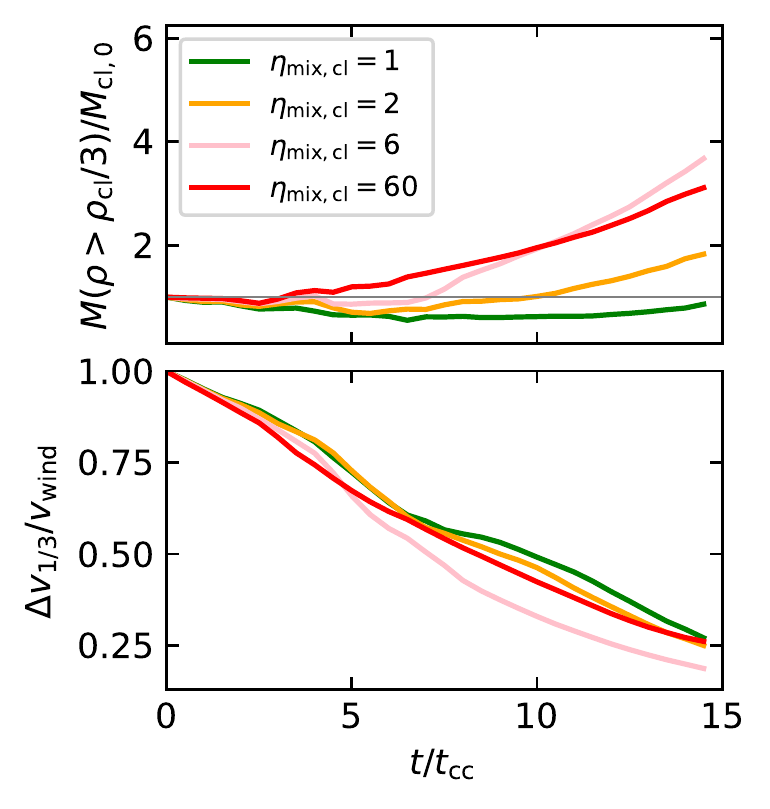}
\caption{\label{figure:broken_plaw_bulk_evo} Like Figure~\ref{figure:bulk_evo} except that the data is for simulations using custom power law cooling function (see Figure~\ref{figure:custom_curves}) with $\chi=100$, $\Mw=1.5$, and $\rcl /\Delta x=8$.
All simulations have the same $\tcoolmix/\tcc$, but show significantly different evolution, demonstrating the importance of cooling below \Tmix.
}
\end{figure}

\begin{figure}
  \center
\includegraphics[width = \columnwidth]{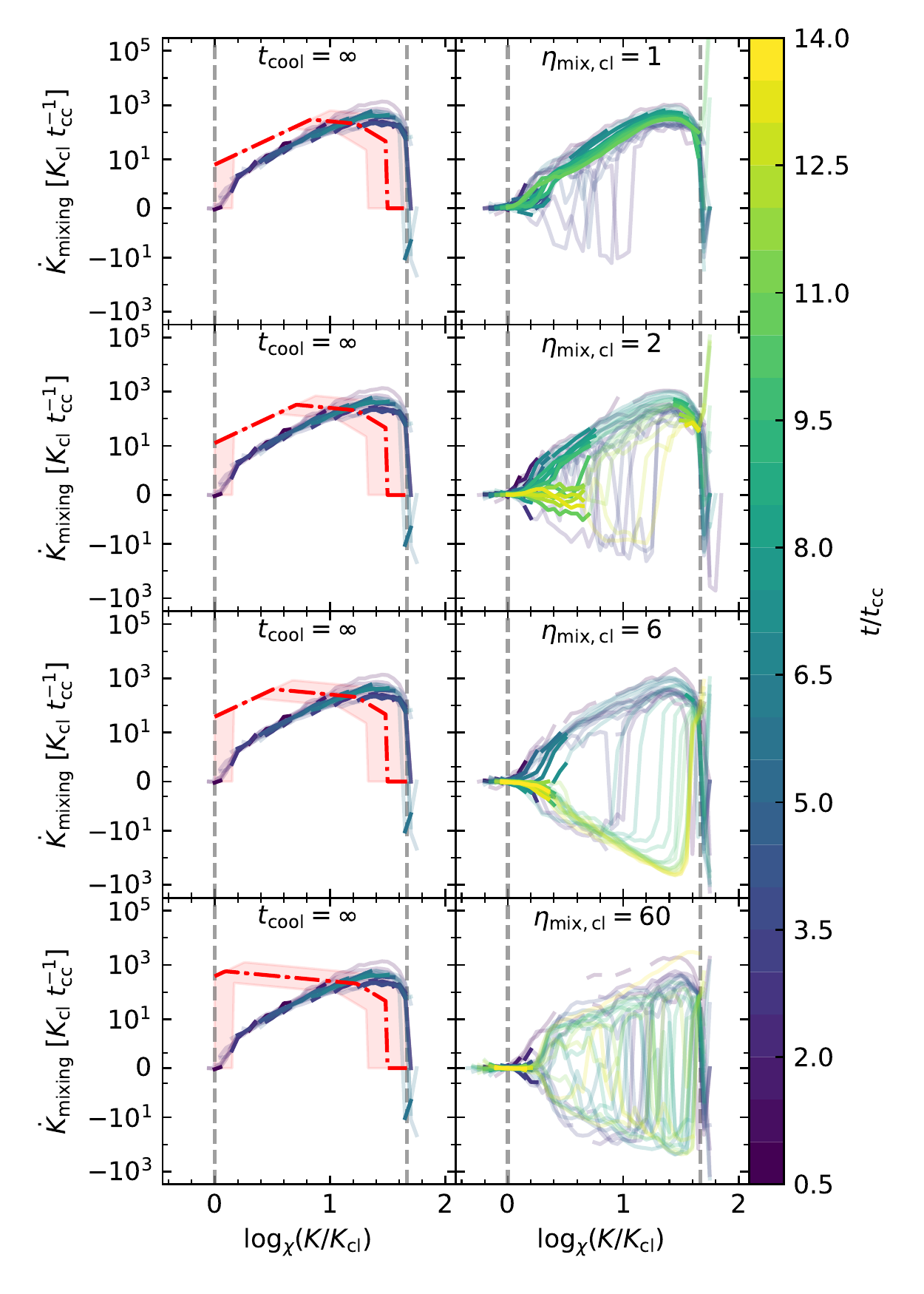}
\caption{\label{figure:dKdt_broken_plaw} Like Figure~\ref{figure:dKdt_cool_adiabat} except that the measured \Kdot\ curves were computed from the histograms shown in Figure~\ref{figure:dMpsdK_custom_curves}. The histograms were measured from simulations using custom power law cooling function (see Figure~\ref{figure:custom_curves}) with $\chi=100$, $\Mw=1.5$, and $\rcl/\Delta x=8$.
}
\end{figure}

Motivated by the impact of cooling curve shape for fixed $\tcoolmix/\tcc$ ratios, in this subsection, we more systematically investigate how simple variations in the local shape of the cooling curves near \Tcl\ (or \Kcl) modify the onset of growth.

To approximately match the realistic cooling curve, but allow us to the change the lower section in a systematic way, we model the cooling-curve with a three-piece broken power law. 
The shape at low $T$ is controlled by a single parameter: $\plawratio=\tcoolmix/\tcoolcl$.
For $\mathcal{T} = T/\Tmix$, the cooling time is given by
\begin{equation}
  \label{eqn:broken_plaw}
  \frac{\tcool(T)}{\tcoolmix} = 
  \begin{cases}
    \plawratio^{-1} & \mathcal{T} \leq \plawratio^{0.5} \\
    \mathcal{T}^2 & \plawratio^{0.5} < \mathcal{T} \leq \alpha^{-5/9} \\
    \alpha \mathcal{T}^{3.8} & \alpha^{-5/9} < \mathcal{T}
  \end{cases}.
\end{equation}
The value of $\alpha$ is 0.138259 and is set by the intersection of the middle and upper segments\footnote{ The upper power law segment mimics properties of the cooling curve used for \simRCFastFour. 
They share the same $(\tcool(T)/\tcoolmix)$ at $\mathcal{T}=10^{0.65}$ and have comparable power-law slopes from there to $\mathcal{T}=10^{0.75}$.}.
For simplicity, we define $\Lambda(T)$ such that the above equation is satisfied at all pressures and for constant $\mu$\footnote{
  Ordinarily, $\mu$ drops by \tsim{21\%} between $10^4\, {\rm K}$ and $4\times 10^4\, {\rm K}$. 
  Above that, it only drops by $\tsim{4\%}$.
  With a modified version of \simRCFastOne, we confirmed that these $\mu$ variations have negligible impact on mass evolution.
  This simulation used $\rcl/\Delta x=8$, a fixed $\mu$, a re-scaled $\Lambda(\rho,T)$ such that $\tcool(\rho,T)$'s shape is unchanged, and a \rcl\ that maintained $\tcoolmix/\tcc = 0.1576$.
}.


We run four simulations which resemble the fast cooling simulations, but use constant $\mu$ and employ broken power law cooling functions with $\plawratio\in [1,2,6,60]$ (\simBPLawOne, \simBPLawTwo, \simBPLawSix, \simBPLawSixty).
For consistency with earlier simulations, the cooling functions are truncated below \Tcl\ and above $0.6\Tw$.
Figure~\ref{figure:custom_curves} illustrates the shapes of the cooling curves normalized by the properties at the mixing layer.

Figure~\ref{figure:dMpsdK_custom_curves} shows the dependence of the late-time passive scalar mass-entropy profiles on \plawratio.
The high \plawratio\ simulations resemble \simRCFastFour, and as \plawratio\ decreases, they begin to more closely resemble \simRCFastOne.
The temporal stability of these profiles indicates that each simulation has a bistable medium.
However, \simBPLawOne's cold phase may not yet be stable; its peak near \Kcl\ (\Kw) decreases (increases) by a factor of \tsim{2} between $11\tcc$ and $14\tcc$.
Note that \simRCFastFour's profiles in Figure~\ref{figure:dMpsdK_cool_adiabat} suggest that for high \plawratio\ cases, there may be more variability at intermediate $K$ at higher resolutions.

Figure~\ref{figure:broken_plaw_bulk_evo} establishes a clear trend: as \plawratio\ increases and the break in the cooling curve moves to lower $T$, rapid cloud growth sets in more quickly.
This figure also supports our expectation that \simBPLawOne\ is just starting to grow in mass between $11\tcc$ and $14.5\tcc$.
The overtaking of the mass and velocity growth in \simBPLawSixty\ by \simBPLawSix\ may be a resolution effect. Figure~\ref{figure:converge_bulk} of Appendix~\ref{appendix:convergence} shows that velocity evolution is particularly sensitive to resolution.

Our results clearly demonstrate that while \tcoolmix\ is important for identifying the conditions under which cooling occurs, the shape of the cooling curve below \Tmix\ is important for determining when rapid growth commences.

To gain some intuition for why these different cooling times matter, we examine the $\Kdotavgtotal(K,t)$ measurements in Figure~\ref{figure:dKdt_broken_plaw}.
Unsurprisingly, these measurements resemble the fast cooling simulations; it's most obvious when considering measurements from simulations of comparable resolution (see Figure~\ref{figure:converge_Kdot} from Appendix~\ref{appendix:convergence}).
As expected, the $\plawratio\leq 2$ cases (when the power-law break is at high $T$) resemble \simRCFastOne, while the $\plawratio\geq 6$ cases resemble \simRCFastFour.
Comparing the relative magnitudes of $\Kdotavgmixing(K,t)$ and $\Kdotcool(p_0,K)$, in the middle left two panels provide some intuition for why there is such a big difference between $\plawratio=2$ and $\plawratio=6$.



However, the precise property of the cooling curve on the interval $\Tcl\la T\la \Tmix$ that controls the growth rate remains somewhat ambiguous. 
We speculate that the crucial quantity is some kind of (weighted) average over the interval, possibly related to (although not exactly equal to) \tcoolcl\ or min \tcool, and hereafter refer to it as the characteristic cooling time of the cold phase \tcoolclchar.
Regardless of \tcoolclchar's true nature, Figure~\ref{figure:tcool} clearly illustrates that it must be considerably smaller for \simRCFastFour\ than it is for \simRCFastOne.
Therefore, rapid growth commences more quickly in \simRCFastFour.

The onset of rapid growth may coincide with the transition between the ``tail growth" and ``entrained phases" of the cloud's areal growth \citep{gronke20a}.
Because areal growth is more rapid in the earlier phase, this transition likely corresponds to the point when the system is able to reach equilibrium and cooling is able to balance the destructive mixing effects \citep{fielding20a}.
If true, then the delayed transition may imply that a larger area is required when \plawratio\ is smaller.
Such an interpretation would be consistent with the mixing layer being linked to both \tcoolclchar\ and \tcoolmix, rather than just the latter.

\section{Discussion}
\label{section:discussion}

\subsection{What does our model offer?}

Our entropy evolution mixing model provides three main benefits. First, it offers a condition-agnostic method for the characterization and quantification of the cloud-wind interaction's evolution.
Our non-radiative parameter study showed that the model meaningfully captures the processes of cloud destruction.
For idealized, hydrodynamic interactions, it reproduces the well-known destruction time-scale \tcc.
However, the independence of these characterizations with respect to the interaction's physical conditions warrants emphasis.


Our model offers a robust approach for describing cloud destruction in circumstances where the \tcc\ description breaks down.
We have already demonstrated that it quantitatively captures the well-documented effects that magnetic draping have on extending the cloud's lifetime \citep[e.g][]{dusri08a,mccourt15a,banda-barragan18a,gronke20a}.
Another interesting application might be characterizing the evolution of networks of small clouds where the idea of having a monolithic cloud with a well-defined \rcl\ does not really apply.

Second, our model facilitates comparisons between the effects of radiative cooling and empirical characterizations of other effects that affect the system's evolution.
We describe a complementary relationship to the $\tcc<\tcoolmix$ survival criterion from \citet{gronke18a} at length in \S\ref{section:mixing_model} and \S\ref{section:entrainment}.
We liken the relationship to that of a distribution function and point estimation; there is a trade off between information content and computational convenience.
In most cases the timescale comparison is sufficient for predicting the system's fate, but our model can be used to build additional insight.

Our case study of interactions that included radiative cooling exemplified this relationship.
While the timescale criterion accurately predicted whether the clouds survived, it did not predict the delay in both entrainment and onset of rapid growth in the fast cooling $\Tcl=10^4\, {\rm K}$ case.
However, our mixing model revealed that these factors are sensitive to the characteristic cooling time of the colder phase \tcoolclchar. 
We defer discussions of this finding's significance to \S\ref{section:discuss_trml}.
Future work should develop a simple criterion encoding this information that either supplements or improves upon the existing survival criterion.

Finally, the model's simplicity makes it extendable.
Given the largely unimodal distribution of $p$ at each value of $K$, the model is conducive to layering additional quantities atop $p-K$ space; one could imagine constructing manifolds in higher dimensional space.
For example, one could supplement $p-K$ space with the wind-aligned velocity \citep[similar to ][]{schneider17a,kanjilal20a}.
This would also connect our model to the established relation between a fluid element's wind-aligned velocity and the fraction of its mass that originated in the hot phase \citep[e.g.][Tonnesen \& Bryan, in prep]{melso19a,schneider20a}.

As mentioned earlier, it would also be useful to consider the entropy flow for fluid elements originating in the hotter phase in addition to the fluid elements from the colder phase.



\subsection{Limitations and Missing physics}

The omission of geometric information may be a limitation of our model.
Consider a non-radiative hydrodynamic simulation with an initially turbulent cloud.
Because turbulent clouds are destroyed faster than spherical clouds \citep[e.g.][]{schneider17a}, one might expect to measure larger $\Kdotmixingchar(K)$ and thus predict stricter conditions for turbulent radiative mixing layer entrainment.
In reality, \citet{gronke20a} showed that turbulent clouds not only survive under the same conditions as spherical clouds, but initially grow faster because they have larger surface areas.
Additionally, it's unclear how well the model captures the evolution of a system in which each phase has different levels of non-thermal pressure support.

We note that these proposed limitations are entirely hypothetical.
Simulations are needed to assess whether there are actually issues in these scenarios.
Regardless, we are unaware of any alternative models with similar predictive power that are devoid of these issues.

This work entirely neglected relevant physical effects like viscosity, conduction, and cosmic rays.
It also didn't consider magnetic fields at the same time as radiative cooling.
Additionally, we artificially prevented cooling below \Tcl, which appears to have a large impact on entrainment 
\citep[c.f. Appendix~\ref{appendix:extra_survival_plots}; ][]{gronke18a}.
Moreover, we only considered idealized initial conditions.
The influence of metallicity variations, different magnetic field configurations, and the presence of turbulence warrant attention in future work.
However, we emphasize that our model is well-equipped for characterizing how each of these conditions modify the conditions for turbulent radiative mixing layer entrainment.

\subsection{Turbulent Radiative Mixing Layer Entrainment}
\label{section:discuss_trml}

Our mixing model is conducive to applications related to cloud survival through turbulent radiative mixing layer entrainment.
It naturally provide a general condition under which this entrainment mechanism is expected (i.e. $\Kdotmixing(K,t) + \Kdotcool(p_0,K)\ll 0$ for a sub-interval of $K\in[\Kcl,\Kw]$, see \S\ref{section:mixing_model}) that is useful for building intuition about the process.
However, this condition doesn't replace the more analytic form of survival criteria presented by \citet{gronke18a} and \citet{li20a}.
Whereas such survival criteria facilitate isolated predictions about cloud survival, our mixing model currently requires empirical measurements of $\Kdotmixing(K,t)$ from non-radiative simulations to predict the interaction's fate.
Therefore, our mixing model complements such criteria, and can be used to help improve them.

\subsubsection{Relevant timescale}

Our results in \S\ref{section:entrainment}--\ref{section:custom_curves} suggest that the most important cooling timescales for turbulent radiative mixing layer entrainment are at temperatures ranging from \Tcl\ through \Tmix.
While the \citet{gronke18a} survival criterion, which is based on \tcoolmix, appears to accurately predict cloud survival, it doesn't fully specify the interaction's evolution.
Specifically, the characteristic cooling time of the colder phase, \tcoolclchar, affects how quickly rapid cloud growth commences (the delay from the start of the simulation appears correlated with $\tcoolclchar/\tcoolmix$).

The delay is significant because it provides additional opportunity for other processes (e.g. externally-driven turbulence in the wind) to destroy the cloud.
This raises a broader point.
Although the distinction between cloud survival and destruction is of primary interest, knowing how close a surviving cloud comes to being destroyed (or how quickly rapid growth commences) would be insightful.
We discuss how the delay in rapid growth may affect the prevalence of turbulent radiative mixing layer entrainment in \S\ref{section:discuss_prevalence}.

There is also direct evidence that \tcoolw, which underlies the \citet{li20a} and \citet{sparre20a} criteria, is not the dominant cooling time-scale.
Figure 3 of \citet{gronke18a} and Figure~\ref{figure:vlct_ppm_bulk_evo_comparison} in Appendix~\ref{appendix:extra_survival_plots} show that switching cooling on and off above ${\sim}0.6\Tw$ has minimal effect on the mass evolution for \simRCFastFour\ and \simRCFastOne, respectively.
The main consequence of the wind cooling is that the system's equilibrium pressure drops (by ${\la}30\%$ for \simRCFastFour\ and \simRCFastOne), which does not appear to be significant.
While this does affect the cooling function, we don't expect it to be significant in most cases (see Appendix~\ref{appendix:extra_survival_plots} for further discussion).

%
%
%

The fact that the difference in \tcoolclchar\ between \simRCFastOne\ and \simRCFastFour\ so efficiently accounts for the variations in mass growth rates reinforces our conclusion that \tcoolw\ is not the dominant time-scale. 
If \tcoolw\ were dominant, we would expect it to explain the difference.

Finally, we address \citet{sparre20a}'s proposed explanation for why \tcoolw\ could be important to turbulent radiative mixing layer entrainment: they suggest that a fluid element's temperature evolution from \Tw\ to \Tcl is rate-limited by an initial cooling phase near \Tw, set by \tcoolw.
While we acknowledge that initial cooling of the wind could possibly make entrainment easier, we expect this to be high-order effect.
In fact, the small impact that switching cooling on and off above $0.6\Tw$ has on the cold phase mass-growth suggests that the temperature change at high $T$ is dominated by mixing.

\subsubsection{Comparison with prior work}



Our results are largely consistent with \citet{gronke18a,gronke20a} and \citet{kanjilal20a}, but they need to be reconciled with those of \citet{li20a} and \citet{sparre20a}.
Most works primarily considered a set of ``typical'' conditions with $\chi\geq100$, $\Tcl=10^4\, {\rm K}$, and an $p/k_B = 10^3\,{\rm K}\, {\rm cm}^{-3}$.\footnote{
There are a few notable exceptions.
\citet{gronke18a,gronke20a} considered clouds with $\Tcl=4\times 10^4\, {\rm K}$.
\citet{li20a} considered a larger range of pressures and $\chi$ values.}
\citet{li20a} and \citet{sparre20a}'s results both support survival criteria that require a larger minimum survival radius under these conditions than the \citet{gronke18a} survival criterion.
We largely attribute this difference to a combination of choices, which include truncation of the cooling function, the simulation box size, and the standards for identifying destroyed clouds.

Under these ``typical'' conditions, every survival criterion implicitly requires that $\tcc\ga\tcoolcl$.
Because growth takes a few \tcc\ to develop, unrestricted cooling will cause clouds to initially contract.
Like \citet{gronke18a,gronke20a} and \citet{kanjilal20a}, we explicitly prevent gas below \Tcl\ from cooling.
In contrast, \citet{sparre20a} allows cooling down to $\Tcl/2 = 5\times10^3\, {\rm K}$.
Because this contraction impedes growth (see Appendix~\ref{appendix:extra_survival_plots}) and might make \citet{sparre20a}'s $\Mw\sim4.5$ simulations susceptible to shattering,\footnote{
For $\chi\ga100$ clouds with $\rcl>\lcool$, \citet{gronke20b} show that tripling the \rhocl's initial density from cooling-driven contraction cause shattering.
They further argue that the shock from oncoming winds with $\Mw\ga1.6$ may produce an equivalent effect (when cooling is prevented below \Tcl).} this difference may help to reconcile our results.
However, \citet{sparre20a}'s inclusion of magnetic fields could plausibly inhibit these effects.
While \citet{li20a} didn't truncate their cooling curve, their results probably aren't strongly affected because they allow their initial conditions to equilibrate before introducing the velocity difference.

\citet{kanjilal20a} highlight two choices made by \citet{li20a} and \citet{sparre20a} that may further help to reconcile our results.
First, \citet{kanjilal20a} argue that \citet{li20a}'s small box-size may 
cause misclassification of growing clouds.
While plausible, we note that \citet{li20a} claimed to have verified their conclusions with longer boxes.
Relatedly, the reflection of shocks off of the transverse boundaries, like we encountered for our non-radiative simulations, may introduce some artificial shock heating in \citet{sparre20a}'s $\chi=1000$ simulations.
However, it remains unclear how significant this artifact is in simulations with cooling.

Second, both studies choose standards for cloud survival that implicitly place requirements on when growth commences.
\citet{li20a} and \citet{sparre20a} identify surviving clouds in cases where cloud growth causes the total cold phase\footnote{They each identify the cold phase with the density threshold $\sqrt{\rhocl\rhow}$ rather than $\rhocl/3$.
  We don't expect small differences in thresholds to make a significant difference.
}
mass to never fall below $10\%$ of the initial mass and to have a positive derivative at $t=12.5\tcc$, respectively.
Thus, they classify clouds differently that survive, but come closer to being destroyed \citep[we expect the differences to be minimal for][]{li20a}.
Furthermore, many of \citet{sparre20a}'s $\Mw=0.5$ and $\Mw=1.5$ 
simulations (their mass evolution is shown in Figure 8 and Appendix B1), could plausibly show growth at later times, which would favor the \citet{gronke18a} criterion.


Finally, we acknowledge the possibility that the inclusion of thermal conduction could simply make it more difficult for turbulent radiative mixing layer entrainment to occur;
\citet{li20a} suggests that efficient conduction could make it more difficult for a mixing layer to form at \Tmix.
While we generally believe that \tcoolmix\ and \tcoolclchar\ are the most important time scales for the process of entrainment, it's plausible that process of thermal conduction could increase the importance of \tcoolw.

\subsubsection{Prevalence}
\label{section:discuss_prevalence}

Our result that rapid cloud growth takes longer to commence at larger $\tcoolclchar/\tcoolmix$ has important implications for the prevalence of turbulent radiative mixing layer entrainment.
This effect is most pertinent for clouds at or near thermal equilibrium, where \tcoolclchar\ is largest.
To give a concrete example with a realistic cooling curve, consider a system at $p = 10^3 k_B\, {\rm K cm}^{-3}$ in which $\Mw=1.5$, $\Tcl\sim6\times 10^{3}\, {\rm K}$ and $\chi\ga180$.\footnote{
We conservatively chose this lower bound on $\chi$ to ensure that the equilibrium pressure drop from cooling above $0.6\Tw$ has no more influence on cloud growth than it does for \simRCFastOne\ (see Appendix~\ref{appendix:extra_survival_plots}).}

%
%
%

For such clouds, the delay in growth provides additional opportunity for destructive processes (like mixing) to destroy the cloud.
Therefore, these clouds require a minimum survival radius that is somewhat larger than the \citet{gronke18a} criterion predicts.
Figure~\ref{figure:extra_survival} illustrates this effect for an idealized, initially laminar wind.
However, this delay may be even more relevant for clouds embedded in winds with turbulence driven by external processes (e.g. supernovae) because mixing may more efficiently destroy clouds\footnote{Note, magnetic fields might directly mitigate this.
\citet{banda-barragan18a} showed that magnetic field have a stabilizing effect on turbulent clouds in a laminar wind.
They might plausibly have a similar impact in a wind with externally-driven turbulence.}
In this scenario, one might predict that turbulent diffusion prevents the formation of a near-continuous tail.
Because the tail makes up a large fraction of the cloud's surface area, its accretion rate would be reduced \citep{gronke20a}.

In their high-resolution starburst-driven galactic wind simulation, \citet{schneider20a} cited external turbulence as a potential explanation for the lack of cloud growth at large radii.
Furthermore, their cooling curve's shape and $10^4\, {\rm K}$ floor \citep{schneider18b} make the delayed growth and entrainment, from large $\tcoolclchar/\tcoolmix$, relevant (as in \simRCFastOne).
We expect that the combination of external turbulence and delay potentially impedes growth near the galaxy.
At larger radii, the hot phase's \tsim{10} times larger pressure than the cold phase (in the simulation) may further exacerbate the effect.
If the intermediate phase also has a somewhat elevated pressure, then $\tcoolclchar/\tcoolmix$ should be larger because $\tcool$ has an inverse dependence on pressure for photo-ionized gas.
Although \citet{gronke20a} showed the rapid growth in 
expanding winds, their model explicitly assumes that the cold and hot phases are in sonic contact.

\section{Conclusion}
\label{section:conclusion}

We have presented an entropy-based formalism for interpreting the cloud-wind interaction's evolution.
The basic premise of the approach is that information about the system's state is encoded in the evolution of its thermodynamic phase space. 
We consider $p-K$ phase space to take advantage of the system's quasi-isobaric nature, and the conservation of a fluid element's specific entropy in the absence of irreversible processes (like shocks, mixing, and heating/cooling).
Thus, in the adiabatic limit, the gas distribution along $K$ is primarily governed by the history of mixing, the dominant cloud destruction process.

We leverage the fact that mixing occurs in $K$-space to introduce an empirical mixing model.
We characterize mixing with the average rate of change in the entropy of fluid element's originating in the cloud, $\Kdotmixing(K,t)$.
From this knowledge, we can define a mixing timescale, $K/\Kdotmixing(K,t)$ as a function of $K$ and $t$.
Additionally, the model provides the capability for making predictions about how radiative cooling will modify adiabatic mixing by facilitating comparisons of $\Kdotcool(p_0,K)= K/\tcool(p_0,K)$, with measurements of $\Kdotmixing(K,t)$.

We have considered two example applications, using \enzoe\ simulations, to demonstrate that this mixing model works as expected and provides useful insight.
We enumerate our four main results below:
\begin{enumerate}
\item The timescale of cloud destruction from adiabatic mixing is well characterized by $K/\Kdotmixing(K,t)$ for most entropy values ranging from the initial value in the cloud, \Kcl, to the initial value in the wind. In fact,  $K/\Kdotmixing(K,t)$ is comparable to \tcc\ for idealized, non-radiative, hydrodynamical interactions.

\item In addition, the model can characterize the change in destruction rate due to other physical processes. 
For example, we have demonstrated that the model reflects the reduction in the cloud destruction rate from the tangling of initially transverse magnetic fields.

\item These characterizations are well suited for comparisons against the effects of radiative cooling. An analogous form of the \citet{gronke18a} survival criterion, $\tcc>\tcoolmix$, can be formulated in terms of this model.

\item We used our model to show that the local shape of the cooling curve can influence the process of cloud entrainment via the rapid cooling of gas that has mixed with the wind. Independent of the cooling time at the mixing layer, variations in the characteristic cooling time of the cold phase can more than double the elapsed time required for clouds to commence rapid growth and become entrained.
\end{enumerate}


\acknowledgments

The authors are grateful to James Bordner, Mike Norman, and the other \enzoe\ developers.
We primarily performed simulations and analysis using the NSF XSEDE facility. GLB acknowledges financial support  from the NSF (grant AST-1615955, OAC-1835509, AST-2006176) and computing support from NSF XSEDE and the Texas Advanced Computing Center (TACC) at the University of Texas at Austin. 
We also acknowledge computing resources from Columbia University's Shared Research Computing Facility project, which is supported by NIH Research Facility Improvement Grant 1G20RR030893-01, and associated funds from the New York State Empire State Development, Division of Science Technology and Innovation (NYSTAR) Contract C090171, both awarded April 15, 2010.

%

\vspace{5mm}


\software{yt \citep{turk11a},  
          Launcher Utility \citep{wilson14},
          }



\appendix

\section{Simulation Robustness}
\label{appendix:extra_survival_plots}

\begin{figure}
  \center
\includegraphics[width = 3.5in]{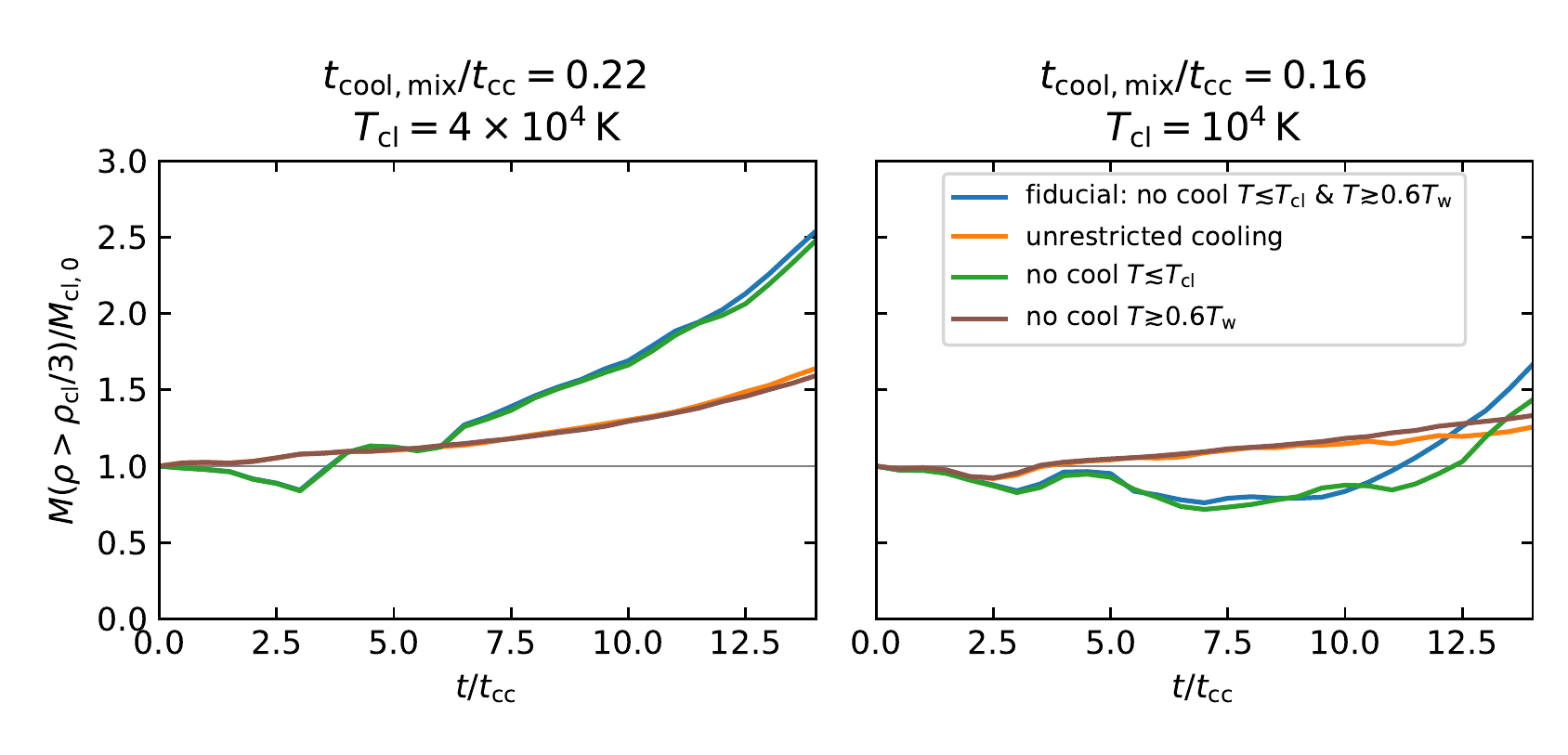}
\caption{\label{figure:bulk_evo_cooling_restrction} Comparison of how cooling curve restrictions affect the mass evolution of the cold phase (gas with $\rho>\rhocl/3$) for low resolution versions ($\rcl/\Delta x=8$)
of both \simRCFastFour\ (left) and \simRCFastOne\ (right).}
\end{figure}

\begin{figure*}
  \center
\includegraphics[width =\textwidth]{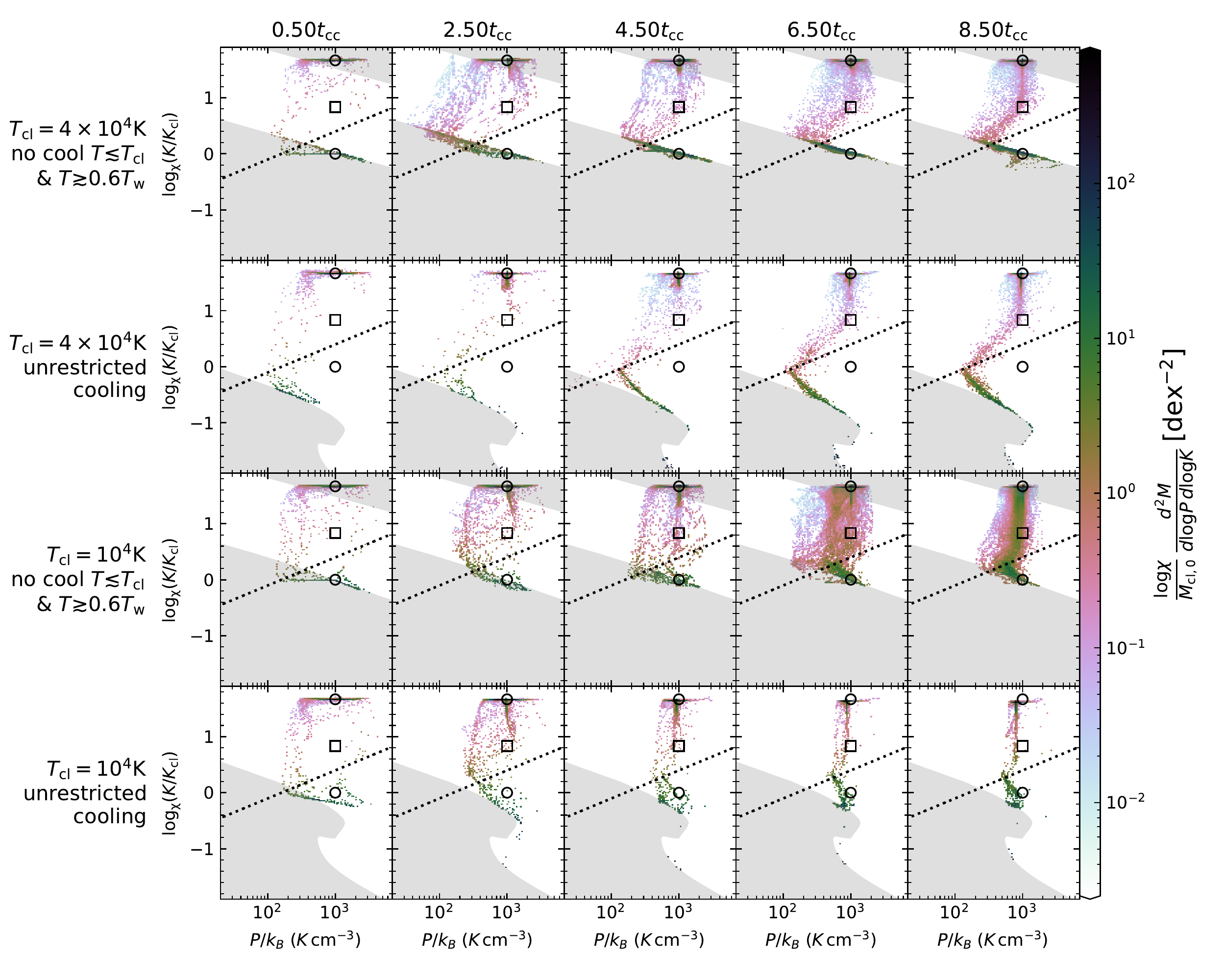}
\caption{\label{figure:pK_restricted_cooling}
  Comparison of how restricting cooling affects the mass weighted pressure-entropy evolution. 
  The top row and second row from the bottom are similar to the bottom two rows of Figure~\ref{figure:pK_cooling}.
  The two differences are that this figure shows the distribution of {\it all} fluid elements in the  simulation domain (not just the fluid elements originating in the cloud) and the simulation resolution is lower ($R_{\rm cl}/\Delta x = 8$).
  The other rows show simulations with same initial conditions, but have unrestricted cooling.
  The gray region indicates wherever $\tcool > 10^4 \tcc$ or heating dominates.
}
\end{figure*}

\subsection{Cooling Curve Restrictions}

Figure~\ref{figure:bulk_evo_cooling_restrction} illustrates how switching cooling on and off at temperatures above \tsim{0.6\Tw} and below \Tcl, affects the cold phase mass evolution for \simRCFastFour\ and \simRCFastOne.
Switching cooling on and off above \tsim{0.6\Tw} is relatively insignificant.
However, when cooling is allowed below \Tcl, the initial period of mass loss is reduced and is followed by a period of slower growth.
These results are consistent with \citet{gronke18a}, who argue that the cooling below \Tcl\ affects mass evolution because it causes the cloud to contract.

Figure~\ref{figure:pK_restricted_cooling} shows how the difference between the fiducial restricted cooling curves (used in the bulk of this work) and the unrestricted cooling curves impact the $p-K$ evolution for {\bf all} gas in \simRCFastFour\ and \simRCFastOne.
There are two main differences.
First, cooling below \Tcl, drives mass to much lower $K$ (or equivalently, $T$).
Second, cooling of the wind slightly decreases (by ${\la}30\%$) the system's equilibrium pressure.

These figures also convey that relaxation of cooling restrictions affect \simRCFastFour\ and \simRCFastOne, when the cooling curves are unrestricted.
Because $\tcoolw/\tcc$ for \simRCFastOne\ is \tsim{46} and is \tsim{3.3} times smaller than for \simRCFastFour, allowing cooling above \tsim{0.6\Tw} causes a slightly larger pressure drop and has slightly more affect on mass growth for \simRCFastOne.
At the same time, \simRCFastFour\ has a significantly larger density increase than  \simRCFastOne; by $t/\tcc=2.5$ the maximum densities have increased to ${\sim}120\rhocl$ and ${\sim}35\rhocl$, respectively.
This makes sense given that the former's \tcc\ is both \tsim{39} times larger than the latter's \tcc\ and ${\ga}100$ times larger than the time needed for gas to cool (whether the cooling proceeds isobarically or isochorically) between the respective \Tcl.

Interestingly, when cooling is unrestricted, \simRCFastFour\ has faster cold phase growth than \simRCFastOne, even though its gas contracts more.
Although increased resolution may affect growth rate, this suggests that the early time value of $\tcoolclchar$ (see \S\ref{section:results-cool}) may be more important for setting the properties of cloud growth.
Future work must investigate cloud evolution using the full cooling curve.

\begin{figure}
  \center
\includegraphics[width = 3.5in]{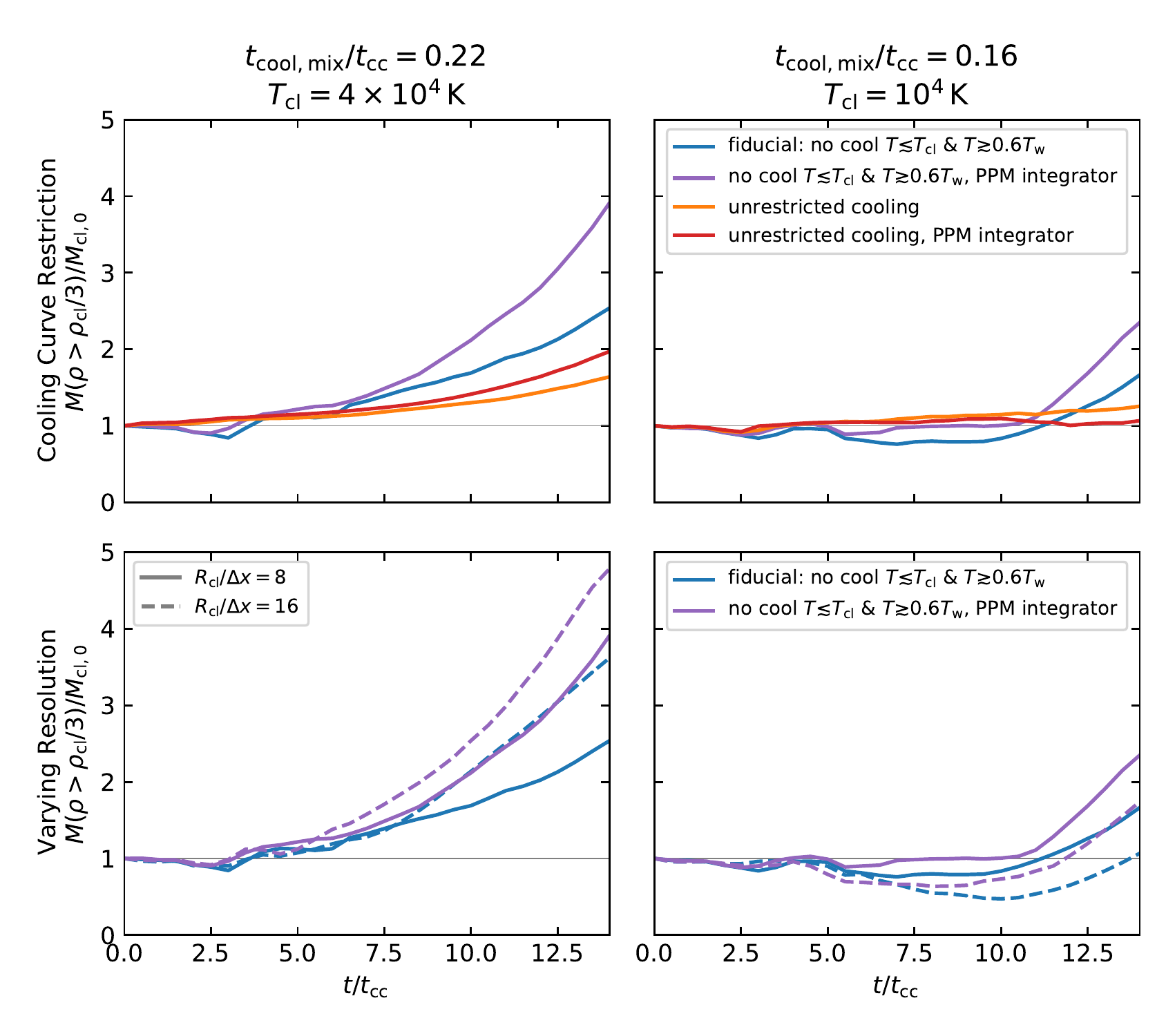}
\caption{\label{figure:vlct_ppm_bulk_evo_comparison} Comparison of how different integrators affect the mass evolution for the cold phase (gas with $\rho>\rhocl/3$) for lower resolution versions of both \simRCFastFour\ (left column) and \simRCFastOne\ (right column). 
The top (bottom) row further shows how the the mass evolution in each integrator is affected by cooling curve restrictions (differences in resolution).
All simulations in the top row have a fixed resolution of $\rcl/\Delta x=8$.
We note that all that the solid orange and blue curves are identical to the curves in Figure~\ref{figure:bulk_evo_cooling_restrction}.}
\end{figure}

\subsection{Influence of Hydrodynamical Integrator}

Figure~\ref{figure:vlct_ppm_bulk_evo_comparison} illustrates how differences in the hydrodynamical integrator modify the cold phase mass evolution.
Specifically we compare the VL+CT integrator (which is used in the rest of this paper) with the ppm integrator, which was previously ported from \enzo\ \citep{bryan14a}. 
The defining differences are that the ppm solver is dimensionally split and employs third order spatial reconstruction. 
Additionally, while the VL+CT integrator employs a predictor-corrector scheme, the ppm solver updates the grid in a single pass.
More minor differences include implementation choices for the dual energy formalism and the choice of the HLLD\footnote{When magnetic fields are zero the HLLD solver reduces to an HLLC solver.
} (Two-Shock) Riemann solver for our simulations with VL+CT (ppm).

We primarily consider how the different integrators affect the mass growth in \simRCFastFour\ and \simRCFastOne\ when using our standard restricted cooling curves (bottom row of Figure~\ref{figure:vlct_ppm_bulk_evo_comparison}).
The simulations using the PPM integrator each show elevated mass growth rates, compared to the VL+CT simulations.
However, we find solace in the way that the simulations using the PPM integrator appear to trend towards the converged curves (discussed in Appendix~\ref{appendix:convergence}) as we increase resolution.

We also compare how the difference in integrators affect the mass growth when cooling is unrestricted (top row of Figure~\ref{figure:vlct_ppm_bulk_evo_comparison}).
Interestingly, the \simRCFastOne\ mass evolution has slower growth when using the PPM curve.
Nevertheless, simulations with both integrators indicate that \simRCFastFour\ shows faster growth than \simRCFastOne.
Our results suggest that while the precise values measured in simulations with different integrators may differ, the trends between the values measured in different simulations (with a single integrator) are fairly robust.

\section{Frame Tracking Scheme Comparison}
\label{appendix:frame_tracking}

\begin{figure*}
  \center
\includegraphics[width =\textwidth]{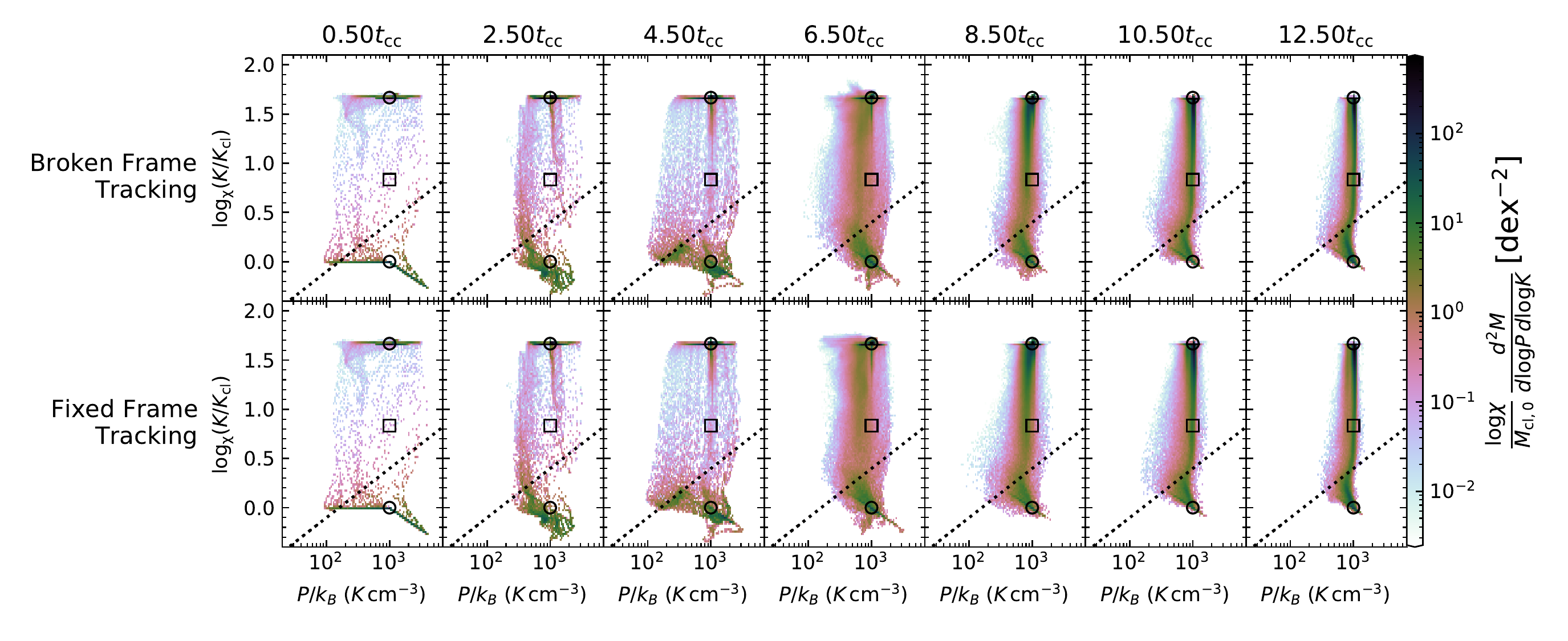}
\caption{\label{figure:pK_cooling_update_comp}
  Comparison of a cloud-wind interaction's mass weighted pressure-entropy evolution simulated with two versions of \enzoe.
  These simulations' initial conditions are the same as those for bottom row of Figure~\ref{figure:pK_cooling}, except that $R_{\rm cl}/\Delta x = 16$.
  Unlike Figure~\ref{figure:pK_cooling}, this figure shows the distribution of {\it all} fluid elements in the simulation domain.
  The simulation in the top row (like all other simulations with $R_{\rm cl}/\Delta x < 64$) effectively has no frame tracking while the bottom row has frame tracking (a slightly refactored Riemann Solver).
  The differences between these simulations are minimal.
}
\end{figure*}

Two different developmental versions of \enzoe\ were employed in this work.
The main difference between them is in the reference frame tracking scheme.
The earlier version (used for simulations with $\rcl/\Delta x < 64$) has a bug which only allows the frame velocity to be updated once, immediately after the very first update cycle.
Thus, the frame velocity remains near zero.

The bug is fixed in the later version (used for simulations with $\rcl/\Delta x = 64$).
Every $0.0625\tcc$, the frame velocity (measured in the cloud's initial reference frame) is updated to match the minimum velocity of cells with a passive scalar density of at least $\rhocl/1000$.
If an update would cause the frame velocity to decrease, it is held constant instead.
This strategy was selected to ensure that the bow shock remained in the simulation domain\footnote{
More aggressive strategies exist that both satisfy this criterion and increase the time that the cold phase remains in the domain}.
The later version of the code also features a more efficient Riemann Solver implementation.

To assess how the code differences affect our results, we compare the results of two simulations using a single set of initial conditions but with the different code versions.
We used initial conditions matching \simRCFastOne, but with a resolution of  $\rcl/\Delta x = 16$.
Figure~\ref{figure:pK_cooling_update_comp} provides a comparison of the phase space evolution for each simulation.
It is clear that both versions of the code produce consistent results.

\section{Resolution Study}
\label{appendix:convergence}
\begin{figure}
  \center
\includegraphics[width = \columnwidth]{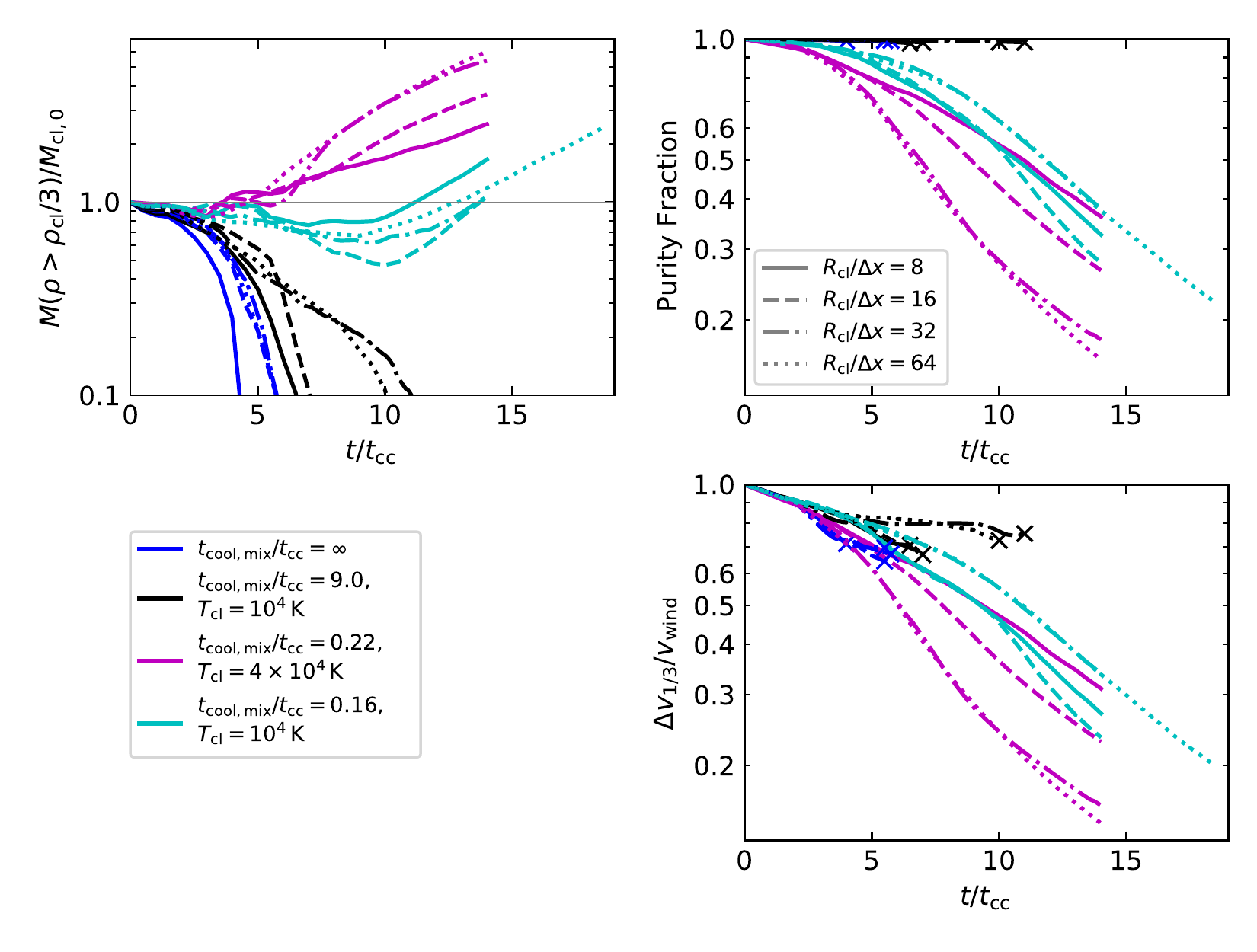}
\caption{\label{figure:converge_bulk} Comparison of how resolution affects the evolution of total mass, purity fraction, and average velocity.
  The evolution clearly converges at high resolution.
}
\end{figure}

\begin{figure}
  \center
\includegraphics[width = 6.5in]{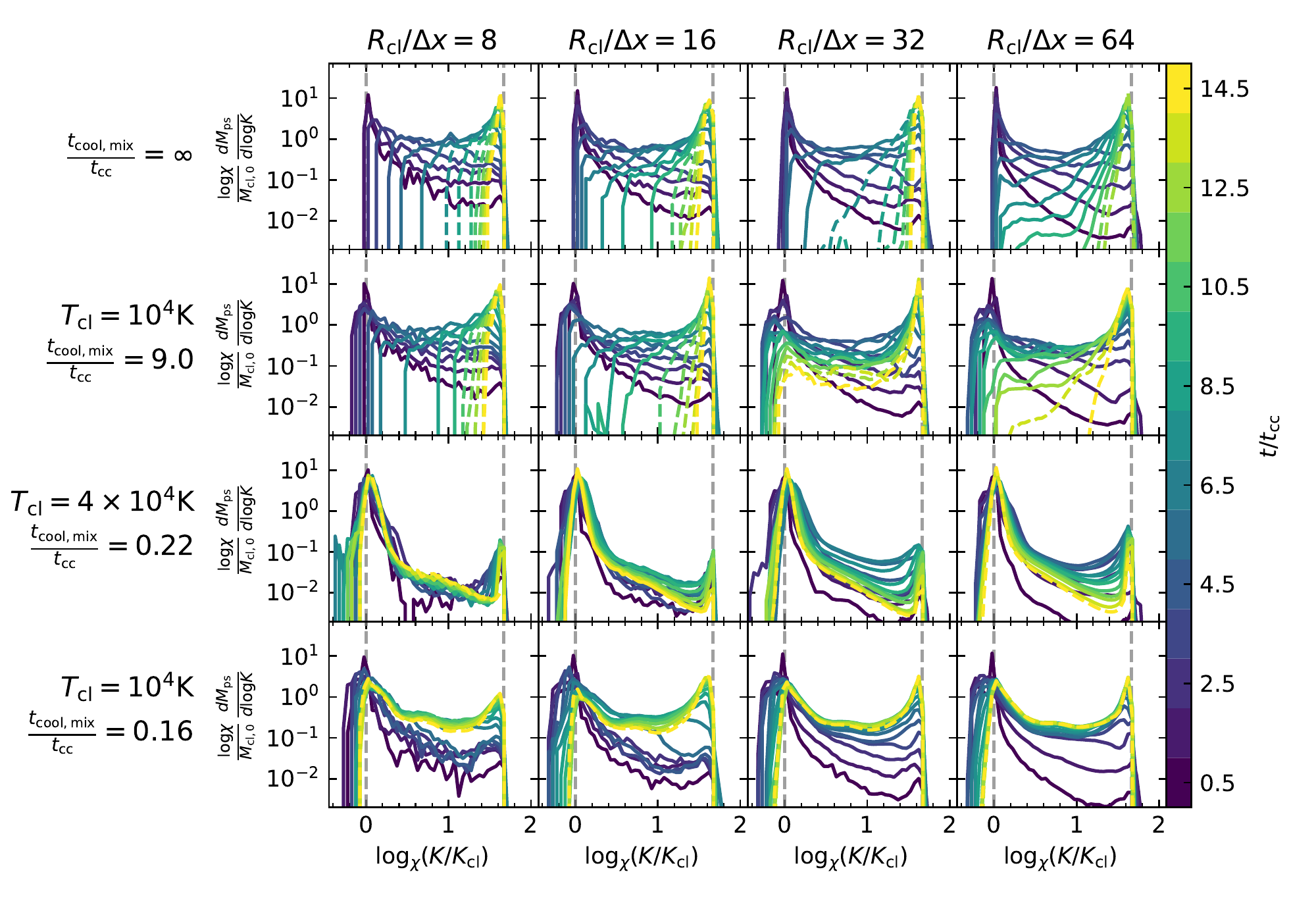}
\caption{\label{figure:converge_dMpsdK} Comparison of $d\Mps/d \log_\chi K$ evolution with respect to resolution. 
Different rows illustrate different simulations and resolution varies between columns.}
\end{figure}

\begin{figure}
  \center
\includegraphics[width = 6.5in]{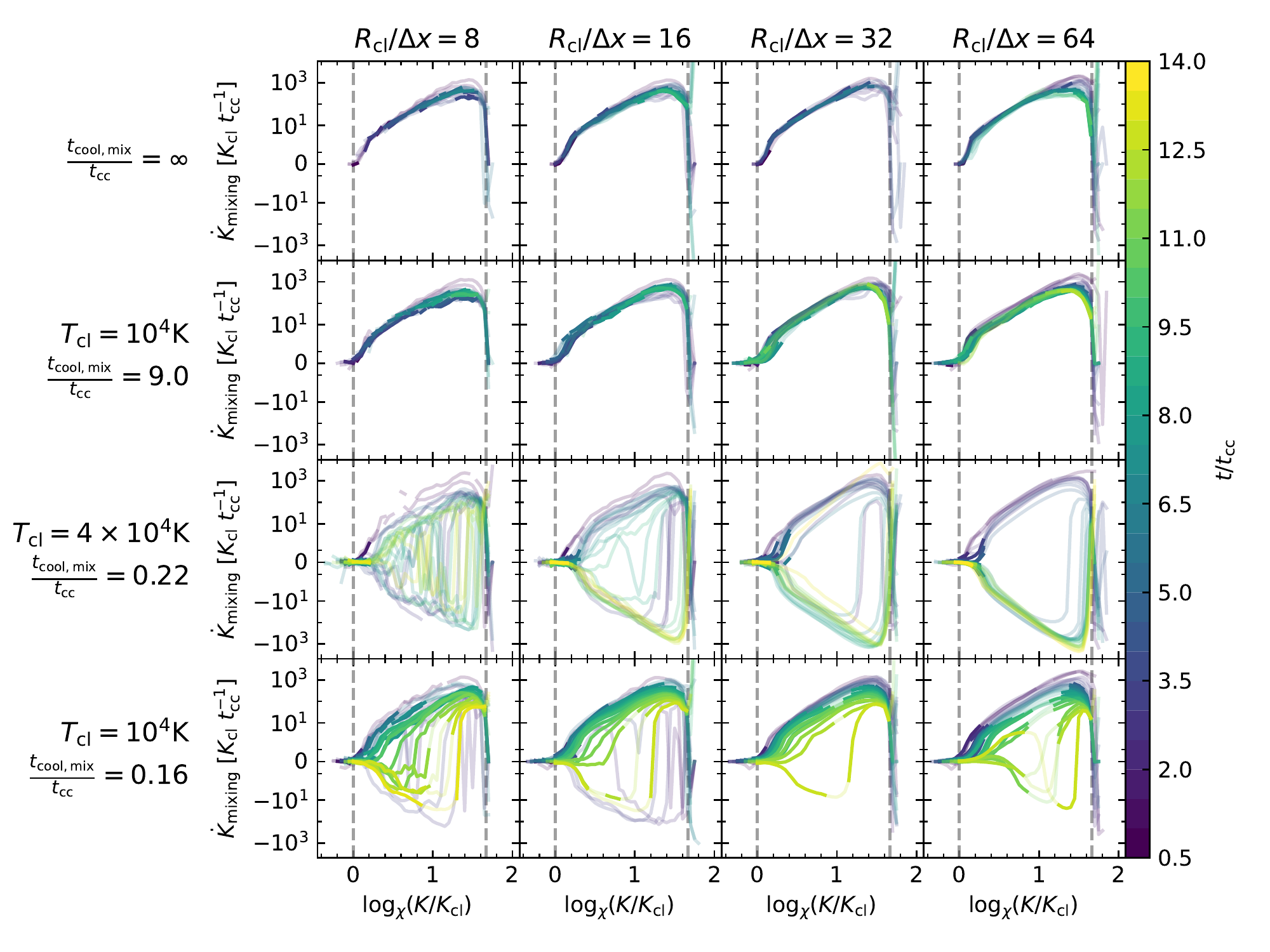}
\caption{\label{figure:converge_Kdot} Comparison of passive scalar mass weighted and time averaged $\Kdot(K)$ at different resolutions.
As in Figure~\ref{figure:converge_dMpsdK}, different rows illustrate different simulations and resolution varies between columns.}
\end{figure}

In this appendix we briefly assess how resolution affects our measurements of the cloud wind-interaction.
To do this we consider the primary four initial conditions discussed in section \S\ref{section:results-cool} (i.e. \simFid, \simRCSlowOne, \simRCFastFour, \simRCFastOne) at the resolutions $\rcl/\Delta x = \{8,16,32,64\}$.

\subsection{Relevance of Shattering}


A relevant length scale for our convergence study is the so-called ``cooling length'', $\lcool \sim \min (c_s t_{\rm cool})$ \citep{mccourt18a}.
\citet{mccourt18a} first showed in 2D simulations that large clouds with sizes exceeding $\lcool$ are prone to fragmenting into a swarm of cloudlets of size \tsim{\lcool}.
Thus, simulations where clouds ``shatter'' may not have well-converged properties when $\lcool$ is not converged.
For reference, Table~\ref{tab:sims} lists each simulation's $\lcool$.
The length scale is well resolved at all resolutions of \simRCSlowOne\ and barely resolved ($\lcool=1.7\Delta x)$ for \simRCFastOne\  when $\rcl/\Delta x = 64$.
However, it's not resolved in any other simulations with cooling.

More recently, \citet{gronke20b} considered cloud shattering in 3D simulations and linked the shattering of clouds to cloud growth through cooling.
They demonstrated that all clouds with $\rcl > \lcool$ that are over pressurized compared to the ambient medium and have density inhomogeneities undergo some degree of shattering.
However, the cloud's fate depends on how the density contrast (when the cloud is over-pressurized), compared to $\chi_{\rm crit}\sim300$.
When the contrast exceeds $\chi_{\rm crit}$, the cloud breaks apart.
Otherwise, the cloud re-coagulates, and has the opportunity to acrete material from the cooling ambient medium.

Although \citet{gronke20b} only studied simulations in which the thermal instability made clouds over-pressurized (the contraction leads to overshooting pressure equilibrium), they argued that similar conditions arise from the shock that supersonic winds drive through clouds.
Because they predict that clouds should only shatter when $\Mw\ga1.6$ (if $\chi=100$ and gas can't cool below \Tcl), we don't expect the clouds in our simulations to shatter.
Nevertheless, resolution of \lcool\ could be important for convergence of simulation properties because of its link to growth.

\subsection{Measurement sensitivity to Resolution}

Figure~\ref{figure:converge_bulk} illustrates how the evolution of the cloud's bulk properties (survival fraction, purity fraction, and bulk velocity) vary with resolution.
The figure illustrates a remarkable level of convergence which seems to imply that the net effects of mixing generally have only a weak dependence on resolution.

Although the net effect of mixing doesn't change significantly, the microscopic details can and do change with resolution.
While we might not expect the average time derivative of $K$ for {\it all} fluid elements to vary much with resolution, the derivative for individual fluid elements can vary wildly.
For this work, we've measured the time-averaged $\Kdotavg(K,t)$ just for fluid elements originating in the cloud.
Therefore, we expect the $\Kdotavg(K,t)$ measurements to be fairly robust for $K$-bins in which the majority of the fluid elements originated in the cloud.
However, when a large fraction of the fluid elements in a bin originated in the wind (and the fluid elements originating from the cloud are no longer representative of all fluid elements in the bin), $\Kdotavg(K,t)$ should be treated with care.

Thus, we expect our $\Kdotavg(K,t)$ measurements for low to intermediate $K$ to be fairly robust for \simFid\ (and non-radiative simulations in general) and \simRCSlowOne\ since the purity fraction remains high.
However, for the fast cooling cases where the purity fraction drops, we expect more variation in $\Kdotavg(K,t)$ at increasing $K$ and over a larger range in $K$ at later times.
These are generally reflected in the convergence properties of \dMpsdKinline\ and \Kdotavg, which are illustrate in Figures \ref{figure:converge_dMpsdK} and \ref{figure:converge_Kdot}.

We note that $\dMpsdKinline(K,t)$ appears to have the most variability in bins with under $1\%$ of the initial mass.
Thus, we focus our assessment of $\Kdotavg(K,t)$ throughout this work on values computed from $K$ bins that include at least $1\%$ of the cloud's initial mass.
Because the measurements that don't satisfy this condition can still be instructive (particularly for simulations with rapid cooling), we still show the other measurements in our figure, as translucent lines.

We further note that calculation of the average \Kdot\ of all fluid elements in the system would improve substantially upon the reliability of our measurements. 
While doing this is possible, we consider our current measurements to be adequate for the purposes of conveying the premise of our mixing model.


\bibliography{manuscript}{}
\bibliographystyle{aasjournal}



\end{document}